\documentclass[pra,aps,onecolumn%,showpacs
]{revtex4}
\usepackage{graphicx}
\usepackage{amsmath}
\usepackage{bbold}
\usepackage{txfonts}
\usepackage{hyperref}
\usepackage{xcolor}
\hypersetup{							    %--- colored links ----
  linkcolor  = red!75!black,
  citecolor  = blue!75!black,
  colorlinks = true
}
\newcommand{\rp}[1]{(\ref{#1})}

\newcommand{\abs}[1]{\left|{#1}\right|}

\newcommand{\av}[1]{\left\langle #1 \right\rangle}

\newcommand{\al}[1]{^{(#1)}}
\newcommand{\da}{^\dagger}

\newcommand{\pt}[1]{\left( #1 \right)}
\newcommand{\pq}[1]{\left[ #1 \right]}
\newcommand{\pg}[1]{\left\{ #1 \right\}}

\newcommand{\lpq}[1]{\left[ #1 \right.}

\newcommand{\rpq}[1]{\left. #1 \right]}

\newcommand{\ee}{{\rm e}}
\newcommand{\ii}{{\rm i}}
\newcommand{\dd}{{\rm d}}

\newcommand{\id}{\mathbb{1}}

\newcommand{\nn}{{\nonumber}}

\newcommand{\mat}[2]{ 
                      \begin{array}{#1}
                       #2
                       \end{array}  }

\newcommand{\va}{{\bf a}}
\newcommand{\vb}{{\bf b}}

\newcommand{\AAA}{{\cal A}}

\newcommand{\CC}{{\cal C}}

\newcommand{\GG}{{\cal G}}

\newcommand{\JJ}{{\cal J}}

\newcommand{\II}{{\cal I}}

\newcommand{\MM}{{\cal M}}

\newcommand{\PP}{{\cal P}}
\newcommand{\QQ}{{\cal Q}}

\newcommand{\TT}{{\cal T}}

\newcommand{\WW}{{\cal W}}

\newcommand{\YY}{{\cal Y}}
\newcommand{\ZZ}{{\cal Z}}

\begin{document}

\title{Entanglement and squeezing of continuous-wave stationary light
}
\author{Stefano Zippilli, Giovanni Di Giuseppe, David Vitali}
\affiliation{School of Science and Technology, Physics Division, University of Camerino, via Madonna delle Carceri, 9, I-62032 Camerino (MC), Italy, and INFN, Sezione di Perugia, Italy}
\date{\today}

\begin{abstract}

Spectral components of continuous squeezed fields are entangled. In this article we review and clarify this phenomenon by analyzing systematically the relations between the correlations of modes filtered from stationary continuous fields and the cross power spectrum between the operators of the corresponding spectral components.
Moreover, we study the specific spectral components that are filtered in homodyne or heterodyne detections and their entanglement properties. In particular, we establish the equivalence between two-mode squeezing variance and logarithmic negativity for the spectral components of continuous stationary fields, thereby demonstrating that the measurement of the homodyne or heterodyne spectrum is, in fact, a direct measurement of the logarithmic negativity between specific spectral modes.
As an illustrative example, we apply these concepts to the analysis of entanglement in ponderomotive squeezing.
\end{abstract}

 
%03.65.-w 	Quantum mechanics
%03.65.Ud 	Entanglement and quantum nonlocality (e.g. EPR paradox, Bell's inequalities, GHZ states, etc.)  
%03.65.Ta 	Foundations of quantum mechanics; measurement theory
%
%03.67.-a 	Quantum information 
%03.67.Bg 	Entanglement production and manipulation
%03.67.Hk 	Quantum communication
%03.67.Mn 	Entanglement measures, witnesses, and other characterizations
%
%42.50.-p
%42.50.Ar 	Photon statistics and coherence theory
%42.50.Dv	Quantum state engineering and measurements in quantum optics
%42.50.Lc	Quantum fluctuations, quantum noise, and quantum jumps
%42.50.Wk	Mechanical effects of light on material media, microstructures and particles
% 
%42.60.-v 	Laser optical systems: design and operation
%42.60.Pk 	Continuous operation
%
%42.65.-k 	Nonlinear optics
%42.65.Lm 	Parametric down conversion and production of entangled photons
%42.65.Yj 	Optical parametric oscillators and amplifiers
%
%85.85.+j 	Micro- and nano-electromechanical systems (MEMS/NEMS) and devices

\maketitle

\section{Introduction}

Quantum optical fields are exploited in the development of a large class of new technologies which make use of quantum mechanics to push their efficiency to the limit~\cite{OBrien}.
In particular, squeezed light plays a pivotal role in the continuous variable domain~\cite{Braunstein,Ralph09,Andersen,Lvovsky14}.
After the first experimental demonstrations of optical squeezing, both in the continuous-wave~\cite{Slusher,Wu} and in the pulsed regime~\cite{Slusher87}, and of the corresponding EPR entanglement~\cite{Ou,Silberhorn,Reid09}, nowadays squeezed optical fields are routinely produced and employed in many experiments aimed at investigating the potentiality of quantum based technologies.
They range, for example, from the demonstration of quantum information tasks such as 
quantum teleportation~\cite{Furusawa98,Takeda}
and other essential elements of scalable universal quantum computation~\cite{Jia,Neergaard-Nielsen,Ukai,Yokoyama,Roslund}, to the design of high-resolution metrology applications~\cite{Xiao,Grangier,LIGO13,Steinlechner,Taylor}, of novel spectroscopic methods ~\cite{Polzik92,Murch}, and of enhanced optical communication schemes~\cite{Slavik}.

Squeezing and entanglement are two very related concepts. In practice squeezed fields are, for example, used to produce two-mode entangled resources by, simply, mixing them on beam splitters~\cite{Silberhorn,Furusawa98,Ukai}. From a more theoretical point of view squeezing variance can be used to construct entanglement criteria~\cite{Reid,Tan,Duan,Giovannetti}.  It is also well known that the spectral components of continuous-wave squeezed light are endowed with non-trivial correlations~\cite{Caves,Yurke,Caves85,Gea-Banacloche87,Ralph08}. 
In particular, specific spectral modes of continuous squeezed fields are entangled~\cite{Schori,Zhang,Huntington,Glockl} realizing EPR spectral beams that have been proposed as convenient quantum communication channels~\cite{Zhang,Hage}.   

In this article we study squeezed continuous fields in the stationary regime, and we analyze the entanglement properties of the corresponding spectral components. We aim at establishing a direct connection between the entanglement theory of continuous-variable systems 
and the spectral properties of squeezed light fields in the stationary continuous-wave regime.
The spectral modes of a continuous field can be operatively defined as the temporal modes filtered from the total field, with a long time filter. 
Their entanglement and squeezing properties are therefore readily defined as the long time limit of 
those that are
found for finite temporal modes.
By employing this approach, we derive general conditions for entanglement and squeezing between two spectral components of stationary continuous fields, and we show that their two-mode squeezing variance can be expressed in terms of the corresponding logarithmic negativity. 
We also discuss the properties of the specific spectral modes that are probed with homodyne and heterodyne detection~\cite{Wu87,Collett}, and we establish that the squeezing spectrum that can be measured with these techniques can be interpreted as a direct measurement of the logarithmic negativity between specific spectral modes.

Finally, we apply these ideas to the analysis of ponderomotive squeezing~\cite{Fabre,Mancini94,Brooks,Safavi-Naeini}, namely the squeezing that is obtained as a result of the optomechanical interaction between a mechanical resonator and the light in an optical cavity.
Optomechanics provides a novel approach to quantum non-linear optics which, recently, has attracted much attention for
its potential applications in quantum enhanced technologies~\cite{Genes,Aspelmeyer}.
In this work, we investigate a two-sided cavity with a membrane in the middle and we identify the spectral components of the output fields that exhibit larger entanglement and that are experimentally accessible with homodyne and heterodyne techniques.

Part of this article comprises a review of already known results, however, rephrased with the intent of providing a clear and complete introduction to the scope of our research. 
In particular, in our presentation, the revision of established concepts is instrumental to the identification and full understanding of the new results concerning the
relationship between entanglement and squeezing in continuous stationary fields, that constitute the central outcome of this article.
In details, the article is organized as follow. In Sec.~\ref{continous} we review the basic properties of continuous fields and of their spectral properties. We introduce the filtered temporal modes and study the correlations of the corresponding field operators in the stationary regime. In Sec.~\ref{squeezEnt}
 we demonstrate the equivalence between logarithmic negativity and two-mode squeezing variance of the spectral components of stationary continuous fields. In Sec.~\ref{homodyne-heterodyne} we review homodyne and heterodyne detection techniques and we study how they can be used to directly measure the logarithmic negativity between spectral modes. Then, in Sec.~\ref{ponderomotive}, we apply the concepts developed in the preceding sections to ponderomotive squeezing. 
Finally, in Sec.~\ref{Conclusions} we draw our conclusions and discuss some possible outlooks.
The three appendices provide additional informations regarding, respectively, the basic properties of entanglement and squeezing of discrete bosonic modes, the 
homodyne and heterodyne techniques, and the input-output theory applied to the investigation of an optomechanical system.

\section{Continuous quantum optical fields}\label{continous}

In this section we introduce the objects of our investigation, namely continuous fields, and we discuss the properties of temporal modes that can be filtered from 
them~\cite{Knoll,Blow,Zhu}. In particular we define the operators for the spectral components in terms of the 
modes that are filtered with a long-time filter and that describe narrow bands of frequencies. These operators are particularly suited for the study of the entanglement  properties of the spectral components of continuous field using standard techniques of entanglement theory.

In detail, we investigate the freely propagating continuous field $E(t)$ at the output of a quantum optical system. It  can be decomposed into the positive and negative frequency components $ E(t)=E\al{+}(t)+ E\al{-}(t)$, with
$ E\al{+}(t)=\ii \int_{0}^\infty\dd \omega\ \sqrt{\frac{\hbar\omega}{4\pi\epsilon_0\,c\,\sigma}}\ \ee^{\ii (k z-\omega t)}\  a(\omega) $
and $E\al{-}(t)=\pg{E\al{+}(t)}\da$, where $\sigma$ is the cross section of the propagating field, and 
$a(\omega)$ is the annihilation operator for the spectral component at frequency $\omega$, which satisfy the standard commutation relation $\pq{a(\omega),\pg{a(\omega')}\da}=\delta(\omega-\omega')$.
In general, in an optical system, only a relatively narrow band of frequencies $\Delta_\omega$ is relevant. This band is centred around the carrier frequency $\omega_L$ of the signal field $E(t)$, which is typically defined by the frequency of a laser driving the system, and fulfils the relation $\Delta_\omega\ll\omega_L$. 
In practice the relevant bandwidth is set by the typical line-width $\Gamma$ of the system under investigation, and eventually by the response time, $T$, of the detector such that $\pg{\Gamma, 1/T}\ll\Delta_\omega$. 
Under these conditions the range of frequency integration in the expression for the field can be extended from $-\infty$ to $\infty$, and the relevant wave numbers $k$ can be approximated with the central value $k\sim\omega_L/c=k_L$.
By this means the quantum optical continuous field can be expressed as
\begin{eqnarray}
E\al{+}(t)=\ii \sqrt{\frac{\hbar\omega_L}{2\,\epsilon_0\,c\,\sigma}}\ee^{\ii k_L z}\ a(t)\ ,
\end{eqnarray}
where we have introduced the continuous field annihilation operator $a(t)$. It is related to the operators for the spectral components by the Fourier transform
$a(t)=\frac{1}{\sqrt{2\pi}}\int_{-\infty}^\infty\dd\omega\,\ee^{-\ii(\omega_L+\omega)t}\,a(\omega_L+\omega)$
where, here, $\omega$ is the frequency relative to the carrier.
It is also useful to introduce the operators 
$\widetilde a(\omega)=a(\omega_L+
\omega)\,\ee^{-\ii\omega_L t}$
relative to the carrier frequency, which are equal to the Fourier transform of $a(t)$
\begin{eqnarray}
\widetilde {{a}}(\omega)&=&\frac{1}{\sqrt{2\pi}}\int_{-\infty}^{\infty} \dd t\ \ee^{\ii\omega t} {{a}}(t)\ ,
\end{eqnarray}
and 
$\pg{\widetilde a(\omega)}\da=\widetilde{\,a\da}(-\omega)$.
These operators satisfy the standard commutation relation for continuous fields
\begin{eqnarray}\label{comm}
&&\pq{\widetilde{a}(\omega),\widetilde{a\da}(\omega')}=\delta\pt{\omega+\omega'}
\nn\\
&&\pq{{a}(t),{a\da}(t')}=\delta\pt{t-t'}\ .
\end{eqnarray}

\subsection{Filtered modes and spectral components of the field}
\label{filteresmodes}

In reality one has access only to a finite time interval (and correspondingly to a  finite band of frequencies) of the total field $a(t)$. These detectable intervals of the total field correspond to specific temporal modes.
They can be physically defined, for example, by the temporal profile of the pumping field in pulsed experiments~\cite{Slusher87,Silberhorn,Smithey,Vasilyev,Hansen}, they can also be extracted by post-processing the previously recorded time signal~\cite{Wasilewski,Kumar}, or they can be selected by the measurement apparatus as a result of the corresponding response and detection times~\cite{Eberly,Yurke85,Breitenbach,Kumar}. In particular, in the case of experiments involving stationary fields, the detection time can be so long to select a well defined spectral component of the total signal (as realized, for example, with an electronic spectrum analyzer)~\cite{Slusher,Wu,Wu87}.

In general a temporal filtered mode can be introduced in terms of a filter function $\phi_\tau(t)$,  which defines the time profile of the mode with a duration of order $\tau$. Correspondingly, it defines a band of spectral components, of width $1/\tau$, that are combined into the filtered signal~\cite{Genes08,Vitali08}.
The generic form for the operators of a filtered mode can be expressed as
\begin{eqnarray}\label{bara}
\overline {a_{\tau}}\pt{\Omega,t}&=&\int_{-\infty}^\infty \dd s\  \ee^{\ii\Omega s}\ \phi_\tau(t-s)\ a(s) 
\end{eqnarray}
with $\pg{\overline {a_{\tau}}\pt{\Omega,t}}\da=\overline {a_{\tau}\da}\pt{-\Omega,t}$,
and where the symbol $\overline{\ \ \vphantom{a}}$ indicates filtered quantities.
The parameter $\Omega$ defines the central frequency of the filter, and the filter function ${\phi_\tau(t)}$ is real and normalized according to
\begin{eqnarray}
\int_{-\infty}^\infty\dd s\ {\phi_\tau(s)}^2 =1\ .
\end{eqnarray}
Consequently, the filtered operators are discrete bosonic operators which satisfy the standard commutation relation $$\pq{\overline {a_\tau}\pt{\Omega,t},{\overline {a_\tau\da}\pt{-\Omega,t}}}=1, \ \ \ \ \ \ \forall\ \tau.$$ 
The corresponding equivalent form of the filtered operators, in terms of the spectral components of the field, is 
\begin{eqnarray}\label{aF2}
\overline {a_\tau}\pt{\Omega,t}&=&\int_{-\infty}^{\infty}\dd\omega\ \ee^{-\ii(\omega-\Omega)\, t}\ \widetilde \phi_\tau(\omega-\Omega)\ \widetilde{a}(\omega)
\end{eqnarray}
where $\widetilde \phi_\tau(\omega)$ is the Fourier transformed filter function
$\widetilde \phi_\tau(\omega)=\frac{1}{\sqrt{2\pi}}\int_{-\infty}^{\infty}\dd s\ \ee^{\ii\omega s} \ \phi_\tau(s),$
that is  peaked at $\omega=0$ and has a width of the order of $1/\tau$.

Two particular cases are worth mentioning. The exponential filter with 
$\phi_\tau^{exp}(t)=\sqrt{2}\theta(t)\ee^{-t/\tau}/\sqrt{\tau}$ , 
and
$\widetilde \phi^{exp}_\tau(\omega)=\sqrt{\tau/\pi}/{\pt{1-\ii\ \tau\ \omega}}$
which has been used, for example, in Ref.~\cite{Eberly} to introduce the physical spectrum of light,
and the step filter function
\begin{eqnarray}\label{phistep}
\phi_\tau^{step}(t)=\frac{\theta(t)\ \theta(\tau-t)}{\sqrt{\tau}}\ , \hspace{0.4cm} \widetilde \phi^{step}_\tau(\omega)=\sqrt{\frac{2\pi }{\tau}}\ee^{\ii \omega\frac{\tau}{2} }\frac{\sin\pt{\omega\ \tau/2}}{\pi\ \omega}
\nn\\
\end{eqnarray}
which we will connect to homodyne and heterodyne detection  techniques in the following.
In this form the time $t$ in the filtered operator $\overline {a_\tau}\pt{\Omega,t}$ corresponds to the final time of the filtering process, that is
$\overline {a_\tau}\pt{\Omega,t}=\int_{-\infty}^t \dd s\  \ee^{\ii\Omega s}\ \phi_\tau(t-s)\ a(s) \ . $
Although not strictly relevant for the results presented in this article, this choice is physically motivated by the fact that in this way we define, at time $t$, a causal operator $\overline a_\tau(\Omega, t)$ which depends only on the past of the continuous field $a(t)$~\cite{Genes08}.

In the limit of long filtering times, $\tau\to\infty$, the filter selects a single spectral component of the field. 
In the following we will focus on the spectral components of stationary fields
for which we will use the following simplified notation
\begin{eqnarray}\label{ainfty}
\overline {a}\pt{\Omega}\equiv\lim_{\tau\to\infty} \overline {a_\tau}\pt{\Omega,t} \ ,
\end{eqnarray}
where we drop the label $\tau$, the limit symbol, and the time argument $t$. 
In particular, the time $t$ in this operators, is irrelevant for stationary fields, in the sense that (as shown in the next section) the correlations of operators of this form, in the limit of large $\tau$, are independent from the time arguments.

We finally note that, we are describing the field in a reference frame rotating at the carrier frequency $\omega_L$, therefore $\overline {a}\pt{\Omega}$ is, in fact, the operator for the spectral component of the field at the sideband frequency $\omega_L+\Omega$.

\subsection{Correlations of filtered spectral modes of stationary fields}\label{correlations}

In this article we are interested in the squeezing properties of the electromagnetic field, which refers to reduced fluctuations or reduced variance of specific quadratures below the vacuum noise level, and in the corresponding entanglement features. Squeezing can be revealed form the analysis of the second order correlations of fields operators. Therefore in this section we analyze the basic properties of the correlations of filtered modes. In particular we focus on the spectral properties of stationary fields. The corresponding field operators, $a(t)$ and $\widetilde a(\omega)$, have diverging correlation functions, as a consequence of the commutation relations in Eq.~\rp{comm}.
In this case is therefore instructive to analyze the fluctuations of the spectral modes in terms of the filtered spectral operators defined in Eq.~\rp{aF2}, whose correlations, on the contrary, are always finite, also in the limit of long integration time $\tau\to\infty$.
This approach is particularly useful for the study of the corresponding entanglement properties, and it has the advantage to provide a clear physical definition of discrete modes corresponding to the specific spectral components, hence allowing for a transparent application of the techniques developed in entanglement theory, which indeed deals with discrete modes (see App.~\ref{discrete}).

Let us study the correlations between the spectral components of two stationary continuous fields with annihilation operators $a_1(t)$ and $a_2(t)$ respectively, which fulfil the commutation relation $\pq{a_j(t),\ {a_k(t')}\da}=\delta_{j,k}\ \delta(t-t')$ (the same results that we discuss below for two continuous fields can be applied with minor modifications to different spectral components belonging to a single field). By straightforward application of the Wiener-Khintchine theorem, we find that all the informations about the correlations between the spectral components are contained into the power spectrum matrix $\widetilde\PP(\omega)$, defined as the Fourier transform of the two-time stationary correlation matrix, 
which can be expressed in terms of the elements of the column vector of operators $\va(t)=\pt{a_1(t),a_2(t),a_1\da(t),a_2\da(t)}^T$, as the matrix $\AAA(t)=\av{\va(t)\,\va(0)^T}$, whose elements are $\pg{\AAA(t)}_{j,k}=\av{\pg{\va(t)}_j\,\pg{\va(0)}_k}$, where $j,k\in\pg{1,2,3,4}$ are vector indices, not to be confused with the indices of the modes. To be specific
\begin{eqnarray}\label{Aom}
\widetilde\PP(\omega)=\int_{-\infty}^\infty\dd t\, \ee^{\ii\,\omega\,\tau}\,\AAA(t)\ ,
\end{eqnarray}
where we use the fact that two-time correlation functions of stationary signals depends only on the difference of the time arguments. 
In particular the correlations between the Fourier transformed operators $\widetilde\va(\omega)=\frac{1}{\sqrt{2\pi}}\int_{-\infty}^{\infty} \dd t\ \ee^{\ii\omega t} {{\va}}(t)$ are diverging and are related to the power spectrum matrix by
\begin{eqnarray}\label{tildeaa}
\av{\widetilde\va(\omega)\, \widetilde\va(\omega')^T}=\delta(\omega+\omega')\, \widetilde\PP(\omega)\ .
\end{eqnarray}
In order to gain insight into the physical meaning of these diverging quantities we employ the narrow filtered modes introduced in the previous section. We construct the vector of filtered spectral components 
$\overline\va(\Omega)=\lim_{\tau\to\infty}\int_{-\infty}^\infty \dd s\  \ee^{\ii\Omega s}\ \phi_\tau(t-s)\ \va(t)$, that is given by $\overline\va(\Omega)=\pt{\overline{a_1}(\Omega),\overline{a_2}(\Omega),\overline{a_1\da}(\Omega),\overline{a_2\da}(\Omega)}^T$,
and we compute the corresponding matrix of correlations
\begin{eqnarray}\label{limAA}
&&\av{\overline\va(\Omega)\, \overline\va(\Omega')^T}=\lim_{\tau\to\infty}
\int_ {-\infty}^{\infty} \dd \omega'\ \int_ {-\infty}^{\infty} \dd \omega\  
\av{\widetilde\va(\omega)\ \widetilde\va(\omega')^T}\
%\nn\\ &&\hspace{2cm}\times\
\ee^{-\ii\pq{(\omega-\Omega) t+(\omega'-\Omega') t'}}\
\widetilde\phi_\tau(\omega-\Omega)\,\widetilde\phi_{\tau}(\omega'-\Omega')\ .\nn\\
\end{eqnarray}
These quantities can be evaluated by noting that for large $\tau$, 
the square modulus of the filter function approaches a delta function, $\lim_{\tau\to\infty}\abs{\widetilde \phi_\tau(\omega)}^2=\delta(\omega)$, while its integral goes to zero,  $\lim_{\tau\to\infty}\int\dd\omega\,\widetilde \phi_\tau(\omega)=0$. And, correspondingly,
given a generic finite function $f(\omega)$, the relation
\begin{eqnarray}\label{rel-filter}
\lim_{\tau\to\infty}\int_{-\infty}^\infty\dd\omega\ \widetilde \phi_\tau(\omega+\Omega)\ \widetilde \phi_\tau(-\omega-\Omega')\ f(\omega)
= \delta_{\Omega,\Omega'}\ f(\Omega)\ 
\nn\\
\end{eqnarray}
holds. 
Consequently Eqs.~\rp{tildeaa} and \rp{rel-filter} can be used in Eq.~\rp{limAA} to find 
\begin{eqnarray}\label{filterA}
\av{\overline\va(\Omega)\, \overline\va(\Omega')^T}=\delta_{\Omega,-\Omega'}\,\widetilde\PP(\Omega)\ ,
\end{eqnarray}
that shows that the power spectrum is directly related to the correlations of narrow filtered modes.
In other terms, in the limit of large integration time $\tau$, i.e. when the bandwidth selected by the filter is sufficiently small, the correlation functions reduce to the power spectrum of the continuous field~\cite{Eberly}. This result is valid when $\tau$ is much larger than the decay time of the signals correlations $\tau_C$ (the memory time of the signals), $\tau_C \ll \tau$. We note in particular that this relation implies the stationarity of the signal which is reached on a time scale of the order of $\tau_C$.

The correlations between two modes are conveniently analyzed in terms of the corresponding correlation matrix, from which the corresponding squeezing and entanglement properties can be readily derived (see App.~\ref{discrete} for a short review). We remark, however, that the matrix $\widetilde\PP(\Omega)$ is not a correlation matrix for two modes. In fact, it contains the correlations between the four spectral modes corresponding to the two pairs of sidebands at frequency $\pm\Omega$ of the two continuous fields. On the other hand, the elements of the power spectrum matrix can be used to construct  the correlation matrix for two narrow modes filtered from the two stationary continuous fields as follows.
We consider  two modes, at frequencies $\Omega$ and $\Omega'$, respectively described 
by the filtered operators 
\begin{eqnarray}\label{filteredoperators}
\overline a_1(\Omega), \ \ \ \ \ \ \overline {a_1\da}(-\Omega), 
\nn\\
\overline a_2(\Omega'), \ \ \ \ \ \ \overline {a_2\da}(-\Omega'),
\end{eqnarray}
where the narrow bandwidth limit $(\tau\to\infty)$ is implicit in their definition [see Eq.~\rp{ainfty}].
The corresponding correlation matrix is defined using the vector 
$\overline{\va}(\Omega,\Omega')=\pt{\overline {a_1}\pt{\Omega}, \overline {a_2}\pt{\Omega'},{\overline {a_1\da}\pt{-\Omega}}, {\overline {a_2\da}\pt{-\Omega'}}}^T$, as
$\overline \AAA(\Omega,\Omega')=\av{\overline{\va}(\Omega,\Omega')\ {\overline{\va}(\Omega,\Omega')}^T}$, and can be expressed in terms of the elements of the power spectrum matrix defined in Eq.~\rp{filterA} as
\begin{widetext}
\begin{eqnarray}\label{AOO}
\overline\AAA(\Omega,\Omega')
=\pt{
\begin{array}{cccc}
\delta_{\Omega,0}\delta_{\Omega',0}\  \pg{\widetilde\PP(0)}_{1,1} &\delta_{\Omega,-\Omega'} \ \pg{\widetilde\PP(\Omega)}_{1,2}& \ \pg{\widetilde\PP(\Omega)}_{1,3} &\delta_{\Omega,\Omega'} \  \pg{\widetilde\PP(\Omega)}_{1,4} \\
 \delta_{\Omega,-\Omega'}  \ \pg{\widetilde\PP(\Omega')}_{2,1} & \delta_{\Omega,0}\delta_{\Omega',0}\  \pg{\widetilde\PP(0)}_{2,2}   & \delta_{\Omega,\Omega'}  \ \pg{\widetilde\PP(\Omega')}_{2,3}&  \ \pg{\widetilde\PP(\Omega')}_{2,4}\\
 \ \pg{\widetilde\PP(-\Omega)}_{3,1} & \delta_{\Omega,\Omega'}  \ \pg{\widetilde\PP(-\Omega)}_{3,2} & \delta_{\Omega,0}\delta_{\Omega',0}\  \pg{\widetilde\PP(0)}_{3,3}  &  \delta_{\Omega,-\Omega'}  \ \pg{\widetilde\PP(-\Omega)}_{3,4}\\
 \delta_{\Omega,\Omega'} \ \pg{\widetilde\PP(-\Omega')}_{4,1} &  \ \pg{\widetilde\PP(-\Omega')}_{4,2} &\delta_{\Omega,-\Omega'} \ \pg{\widetilde\PP(-\Omega')}_{4,3}&\delta_{\Omega,0}\delta_{\Omega',0}\  \pg{\widetilde\PP(0)}_{4,4}    
\end{array}
}\ .
\end{eqnarray}
\end{widetext}
We finally note that 
$\overline\AAA(\Omega,\Omega')$ is equal to the power spectrum matrix $\widetilde\PP(\Omega)$ only at zero frequency,  $\overline\AAA(0,0)=\widetilde\PP(0)$.

\section{
Equivalence between two-mode squeezing variance and logarithmic negativity of the spectral components of  stationary continuous fields}\label{squeezEnt}

Having, in the previous section, introduced our notation, and reviewed the basic properties of stationary continuous fields, we are now in the position to study the general conditions for the squeezing and the entanglement between two spectral components, that can be inferred from Eq.~\rp{AOO}.
In particular, in the case of Gaussian states, we establish the equivalence between the logarithmic negativity and the two-mode squeezing variance of two spectral modes. 

In general, given two modes described by the operators $a_1$ and $a_2$, 
two-mode squeezing is characterized by non vanishing correlations of the form $\av{a_1\,a_2}$ and $\av{a_1\da\,a_2\da}$. According to Eq~\rp{AOO}, the correlation between the annihilation operators for two filtered spectral components of stationary fields, $\overline a_1(\Omega)$ and $\overline a_2(\Omega')$, can be non-vanishing only for opposite frequencies, that is when $\Omega=-\Omega'$. In this case the matrix in Eq.~\rp{AOO} reduces to the form
\begin{eqnarray}\label{Ainfty3}
\overline\AAA(n_+,n_-,m)=
\pt{
\begin{array}{cccc}
0 & m&  n_++1& 0 \\
m & 0   &0 &   n_-+1\\
  n_+& 0 &0 &  m^*\\
 0&  n_- &m^*&0
 \end{array}
}
\end{eqnarray}
where 
\begin{eqnarray}\label{nnm}
n_+&=&\pg{\overline\PP(-\Omega)}_{3,1}\ =\ \av{\overline{a_1\da}(-\Omega)\, \overline{a_1}(\Omega)}
\nn\\
n_-&=&\pg{\overline\PP(\Omega)}_{4,2}\ =\ \av{\overline{a_2\da}(\Omega)\, \overline{a_2}(-\Omega)}
\nn\\
m&=&\pg{\overline\PP(\Omega)}_{1,2}\ =\ \av{\overline{a_1}(\Omega)\, \overline{a_2}(-\Omega)}
\end{eqnarray}
with $n_\pm$ real and positive. Here we have used the general properties of the power spectrum matrix $\widetilde\PP(\Omega)-\widetilde\PP(-\Omega)^T=\pt{\scriptsize\mat{cc}{&\id_2\\ -\id_2&}}
\ $,  where $\id_2$ is the $2\times2$ identity matrix and the missing blocks are null matrices, and $\pg{\widetilde\PP(\Omega)}_{1,2}^*=\pg{\widetilde\PP(-\Omega)}_{3,4}$, i.e. $\av{\overline{a}_1(\Omega)\, \overline{a}_2(-\Omega)}^*=\av{\overline{a_1\da}(-\Omega)\, \overline{a_2\da}(\Omega)}$.
Eq.~\rp{Ainfty3} represent the general form for the correlation matrix between two spectral components at opposite sideband frequencies of stationary continuous fields. This matrix can be
exploited to derive general results regarding the corresponding squeezing and entanglement properties.

In general squeezing refers to the reduced fluctuations of field quadratures.  Let us therefore define the quadrature operators for a spectral mode
\begin{eqnarray}\label{overlineX}
\overline{x_j\al{\theta}}(\Omega)=\ee^{\ii\theta}\,\overline{a_j}(\Omega)+\ee^{-\ii\theta}\,\overline{a_j\da}(-\Omega)\ .
\end{eqnarray}
Thereby, we can formalize the condition for two-mode squeezing of the two spectral components as follows. The two components are 
two-mode squeezed when the variance of a generic composite quadrature of the form 
\begin{eqnarray}\label{Zxi0}
X\al{\theta_+,\theta_-}(\Omega)=\frac{1}{\sqrt{2}}\pq{\overline{x_1\al{\theta_+}}(\Omega) +\overline{x_2\al{\theta_-}}(-\Omega)   }
\end{eqnarray}
is below the shot noise level for some value of $\theta_\pm$ (see App.~\ref{discrete} for further remarks).  
We first note that, in the case of a stationary field, for which $\av{a(t)}=\alpha$  is constant, the average of a corresponding filtered mode is given by $\av{\overline{a_\tau}(\Omega,t)}=\sqrt{2\pi}\,\alpha\,\ee^{\ii\,\Omega\,t}\,\widetilde\phi_\tau(-\Omega)$, and it approaches zero for large $\tau$ and for non-zero values of $\Omega$.
Therefore, according to our definitions, the fields are two-mode squeezed if for some values of $\theta_\pm$ the autocorrelation function of the combined quadrature 
$$\Delta X\al{\theta_+,\theta_-}(\Omega)=
\av{\pq{X\al{\theta_+,\theta_-}(\Omega)}^2}$$
is smaller than one, 
$
\Delta X\al{\theta_\pm}(\Omega)<1$. 
More in general one can construct composite quadratures with different weights of the two components
\begin{eqnarray}\label{Zxi}
X_{\xi_+,\xi_-}\al{\theta_+,\theta_-}(\Omega)=\frac{1}{\sqrt{\xi_+^2+\xi_-^2}}\pq{\xi_+\, \overline{x_1\al{\theta_+}}(\Omega) +\xi_-\,  \overline{x_2\al{\theta_-}}(-\Omega)   }\ .
\end{eqnarray}
It has been shown~\cite{Duan} that the variance of quadratures of this form can be used to define 
entanglement criteria. Specifically when the relation 
\begin{eqnarray}\label{entcriterion}
\Delta X_{\xi_+,\xi_-}\al{\theta_+,\theta_-}(\Omega)\ +\ \Delta X_{\xi_+,\xi_-}\al{\theta_++\frac{\pi}{2},\theta_--\frac{\pi}{2}}(\Omega)<2
\end{eqnarray}
is satisfied for some values of $\theta_\pm$ and $\xi_\pm$, then the two modes are entangled. In general this is a sufficient condition for entanglement, but it is also necessary in the case of Gaussian fields and for an appropriate choice of $\xi_\pm$.
The calculation of the autocorrelation function of these composite quadratures is straightforward using the matrix of correlations in Eq.~\rp{Ainfty3}. 
The result is 
\begin{eqnarray}\label{DeltaXxi}
&&
\Delta X_{\xi_+,\xi_-}\al{\theta_+,\theta_-}(\Omega)=
%\\&&\hspace{0.cm}=
1+
\frac{2 n_+\,\xi_+^2+2 n_-\,\xi_-^2+
2\xi_+\,\xi_-\,\pq{m\,\ee^{\ii(\theta_++\theta_-)}+m^*\,\ee^{-\ii(\theta_++\theta_-)}}
}
{\xi_+^2+\xi_-^2} \ .
\nn
\end{eqnarray}
In particular, we find that $\Delta X_{\xi_+,\xi_-}\al{\theta_+,\theta_-}(\Omega)\ =\ \Delta X_{\xi_+,\xi_-}\al{\theta_++\frac{\pi}{2},\theta_--\frac{\pi}{2}}(\Omega)$, hence, in our case, the condition for entanglement reduces to $\Delta X_{\xi_+,\xi_-}\al{\theta_+,\theta_-}(\Omega)<1$.
The corresponding optimized squeezing spectrum can be defined as the minimum of this quantity over 
the quadrature of the field. Specifically we can identify two different minimization strategy. If we restrict to composite quadrature that are symmetric superposition of the two components ($\xi_+=\xi_-$) as in Eq.~\rp{Zxi0} then the minimization runs only over the phases $\theta_\pm$, and the corresponding phase-optimized squeezing spectrum takes the general form 
\begin{eqnarray}\label{S}
S(\Omega)&=&{\rm min}_{\theta_\pm}\ \Delta X\al{\theta_+,\theta_-}(\Omega)
\nn\\&=&1+n_++n_--2\,\abs{m}\ .
\end{eqnarray}
This is the quantity that is obtained, for example, by the homodyne measurement of a continuous field, where the phases $\theta_\pm$, in Eq.~\rp{DeltaXxi}, are directly related to the phase of the local oscillator (see Sec.~\ref{homodyne-heterodyne} for further details). If, on the other hand, we consider the more general quadratures with the two components scaled by factors $\xi_\pm$ as in Eq.~\rp{Zxi}, then the minimization can be performed both over the phases and over the parameters $\xi_\pm$, and the corresponding globally-optimized squeezing spectrum reduces to 
\begin{eqnarray}\label{Smin}
S_{\rm min}(\Omega)&=&{\rm min}_{\theta_\pm,\xi_\pm}\ \Delta X_{\xi_+,\xi_-}\al{\theta_+,\theta_-}
\nn\\&=&1+n_++n_--\sqrt{4{\abs{m}}^2+\pt{n_+-n_-}^2}\ .
\end{eqnarray}
In general $S_{\rm min}(\Omega)\leq S(\Omega)$, and they are equal in the case of symmetric spectral components, for which $n_+=n_-$.
In the next section we will describe how to measure both quantities in few specific cases with homodyne and heterodyne techniques. Here we emphasize that the occurrence of $S_{\rm min}(\Omega)<1$ implies that the entanglement criterion in Eq.~\rp{entcriterion} is satisfied and, in turn, it entails that {\it squeezing spectrum smaller than one is always a signature of entanglement}. We further note that Eq.~\rp{Smin} is, indeed, smaller than one if and only if
\begin{eqnarray}\label{Ennm}
n_+\,n_-<\abs{m}^2\ .
\end{eqnarray}
Consequently this relation can be interpreted as a sufficient condition for the entanglement between the spectral components at opposite sideband frequencies of stationary continuous fields. And, as already noted, it is also a necessary condition  in the case of Gaussian fields.

Let us now focus to the Gaussian regime. In this case the logarithmic negativity, that is a measure of bipartite entanglement~\cite{Vidal02}, can be expressed as 
\begin{eqnarray}\label{logneg0}
E_N={\rm min}\pg{0,-\log_2(\nu)}
\end{eqnarray}
where the parameter $\nu$ is equal to the smallest symplectic eigenvalues of the covariance matrix corresponding to the partially transposed state of the two modes (see App.~\ref{discrete}). Using Eq.~\rp{Ainfty3} we find that the parameter $\nu$ evaluated for each pair of spectral components at the sideband frequencies $\pm\Omega$ is equal to the squeezing spectrum in Eq.~\rp{Smin},
\begin{eqnarray}\label{nuOmega}
\nu(\Omega)=S_{\rm min}(\Omega) \ .
\end{eqnarray}
This is a general result valid for the spectral components of stationary continuous fields (a specific example has been discussed in Ref.~\cite{Zippilli14}). In particular, this relation implies that, {\it if the two stationary fields are Gaussian, then  a measure of the minimum variance of a composite quadrature of two spectral components, of the form of Eq.~\rp{Zxi}, is a direct measurement of the corresponding logarithmic negativity}.

\section{Homodyne and heterodyne detection of the spectral components of stationary continuous fields}\label{homodyne-heterodyne}

The quadratures of continuous electromagnetic fields are routinely measured in experiments with homodyne and heterodyne techniques~\cite{Collett,Lvovsky,Yuen78,Yuen83,Lvovsky14,Barbosa}. The  photocurrents resulting from homodyne and heterodyne detections are, in fact, proportional to specific quadratures of the detected field. In turn, the power spectrum of the photocurrent, namely the homodyne or heterodyne spectrum, measures the fluctuations of the quadratures at specific frequencies. Such spectra are therefore directly related to the squeezing and entanglement properties of the spectral components of the electromagnetic field, and in particular to the squeezing spectra defined in Eqs.~\rp{S} and \rp{Smin}. Specifically,  we will show that the autocorrelation function of the photocurrent minimized over experimentally accessible parameters as for example the phase of the local oscillator can be always cast in the form of Eqs.~\rp{S} or \rp{Smin}, with corresponding parameters $n_\pm$ and $m$ evaluated for specific spectral modes.  This justifies the interpretation of the optimized homodyne and heterodyne spectra as a direct measurement of the  logarithmic negativity of these modes.

\subsection{Single-mode homodyne spectrum and entangled spectral components}\label{singelhomodyne}

\begin{figure}[!t]
\begin{center}
\includegraphics[width=7cm]{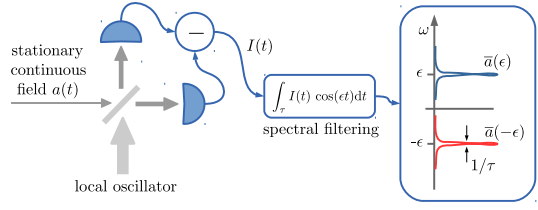}
\end{center}
\caption{Single-mode homodyne detection: A stationary continuous field is detected by homodyne techniques and the 
resulting filtered 
photocurrent is a superposition of spectral components at opposite sideband frequencies $\pm\epsilon$.
}\label{fig1}
\end{figure}

In homodyne detection the signal field is mixed on a 50:50 beam splitter with a strong monochromatic field (the local oscillator) at the same frequency of the carrier signal. The fields at the two output ports of the beam splitter are detected and the corresponding photocurrents are subtracted resulting in a signal which contains informations about a field quadrature~\cite{Yuen83}, and that can be described by a photocurrent operator of the form (see App.~\ref{apphomodyne-heterodyne})
\begin{eqnarray}\label{It0}
I\al{\theta}(t)= \ee^{\ii\theta}\ a(t) +\ \ee^{-\ii\theta}\ a\da(t) \ ,
\end{eqnarray}
where $\theta$ is the phase of the local oscillator. 
The power spectrum of the photocurrent contains informations about the spectral components of the detected field, and  in particular it quantifies the strength of the fluctuations at specific frequencies. We will refer to it as the homodyne spectrum, and it can be expressed as the autocorrelation function of the filtered photocurrent, integrated over a long time $\tau$, of the form $ {J\al{\theta,\varphi}_{\tau}}(\epsilon,t)\propto
\frac{1}{\sqrt{\tau}}\int_{t-\tau}^t \dd s\   \cos(\epsilon\,s+\varphi)\ I\al{\theta}(s)$. In detail, the homodyne spectrum can be written as
\begin{eqnarray}\label{GG}
\GG\al{\theta}(\epsilon)=\lim_{\tau\to\infty}\av{\pq{ {J\al{\theta,\varphi}_{\tau}}(\epsilon,t)}^2}\ .
\end{eqnarray}
We note that, for stationary processes, this quantity is independent from the phase of the filter $\varphi$ (see App.~\ref{apphomodyne-heterodyne}). However, this phase is relevant and can be useful when considering combinations of filtered photocurrents at different phases which, as discussed below, can be exploited to probe arbitrary superpositions of spectral modes. Moreover the same results for the power spectrum in Eq.~\rp{GG} is obtained when, in the filtered photocurrent $J_\tau\al{\theta,\varphi}(\epsilon, t)$, one uses an exponential oscillating function, as in the filters of Sec.~\ref{filteresmodes}, in place of the sinusoidal function introduced above. The use of a sinusoidal function is, however, more convenient because, in this case, the filtered photocurrent is an hermitian operator thereby making the relation between the photocurrent and the field observables more transparent. Specifically the filtered photocurrent in the limit of long filtering time can be expressed as the sum of two filtered quadrature operators for the 
two spectral modes at frequencies  $\pm\epsilon$ (see App.~\ref{apphomodyne-heterodyne})
\begin{eqnarray}\label{narrowI}
 {J\al{\theta,\varphi}}(\epsilon)&=&\lim_{\tau\to\infty} {J_{\tau}\al{\theta,\varphi}}(\epsilon,t)
\nn\\
&=&\frac{1}{\sqrt{2}}\pq{\overline{x\al{\theta+\varphi}}(\epsilon)+ \overline{x\al{\theta-\varphi}}(-\epsilon)}
\ .
\end{eqnarray}
Similarly to Eq.~\rp{Zxi0}, it is a symmetric superposition of two quadratures, defined as in Eq.~\rp{overlineX}, corresponding to the two spectral components, which in this case are filtered from the same field,
and whose annihilation operators are
\begin{eqnarray}\label{aepsilon1}
\overline a(\epsilon) \ \ {\rm and} \  \ \  \overline{a}(-\epsilon)\ ,
\end{eqnarray}
as illustrated in Fig.~\ref{fig1}. The corresponding single-mode homodyne spectrum (where single-mode indicates that it results form the detection of a single continuous field) is equal to Eq.~\rp{DeltaXxi} with $\xi_+=\xi_-$ and $\theta_++\theta_-=2\theta$, 
i.e. $\GG\al{\theta}(\epsilon)=1+n_+\al{I}+ n_-\al{I}+\pq{m\al{I}\,\ee^{2\,\ii\,\theta}+{m\al{I}}^*\,\ee^{-2\,\ii\,\theta}}$,
where 
\begin{eqnarray}\label{nnm1}
n\al{I}_\pm=\av{\overline{a\da}(\mp\epsilon)\, \overline{a}(\pm\epsilon)},\ \  \ \ \  \ \ m\al{I}=\av{\overline{a}(\epsilon)\, \overline{a}(-\epsilon)}\ .
\end{eqnarray}
Thus, the single-mode phase-optimized squeezing spectrum, that is experimentally accessible by tuning the phase of the local oscillator, $S\al{I}(\epsilon)={\rm min}_\theta\ \GG\al{\theta}(\epsilon)$, 
is equal to Eq.~\rp{S} evaluated for the parameters in Eq.~\rp{nnm1}. When it is smaller than one, it indicates that the two sideband modes $\overline a(\pm\epsilon)$ are entangled~\cite{Zhang,Huntington,Glockl,Hage}. 
Their logarithmic negativity, that as discussed in Sec.~\ref{squeezEnt} is directly related to the squeezing spectrum in Eq.~\rp{Smin}, could be measurable if one could construct a filtered photocurrent similar to Eq.~\rp{Zxi}, which is a  non-symmetric superposition of the quadratures of the two modes. This photocurrent is, in fact, realizable by combining two filtered photocurrents detected at 
appropriately tuned phases of the local oscillator $\theta$, and of the filter $\varphi$. 
Specifically one should first detect the filtered photocurrent $ {J\al{\theta,\varphi}}(\epsilon)$ for some value of $\theta$ and $\varphi$, and then a second one  $ {J\al{\theta',\varphi'}}(\epsilon)$, with the phases  tuned  to different values $\theta'$ and $\varphi'$.
 The two photocurrents are then summed resulting in 
the total photocurrent
\begin{eqnarray}\label{Ixi}
 {J_{\xi_+,\xi_-}\al{\theta_+,\theta_-}}(\epsilon)
&=& \frac{1}{\sqrt{\xi_{+}^2+\xi_{-}^2}}\pq{\xi_{+} \overline{x\al{\theta_{+}}}(\epsilon)+ \xi_{-}\   \overline{x\al{\theta_{-}}}( -\epsilon)}\ ,
\end{eqnarray}
that we have appropriately normalized, and where
\begin{eqnarray}\label{tetaxi}
\theta_{\pm}&=&\frac{\theta+\theta'\pm(\varphi+\varphi')}{2}
\nn\\
\xi_{\pm}&=&\cos\pq{\frac{\theta-\theta'\pm(\varphi-\varphi')}{2}} \ .
\end{eqnarray}
We note that Eq.~\rp{Ixi} has, indeed, the form of the composite quadrature defined in Eq.~\rp{Zxi}.
%The combination of filtered photocurrents at different filter phases can be experimentally realized, for example, by recording the photocurrent for a sufficiently long time and then post processing the recorded signal.
The corresponding single-mode globally-optimized squeezing spectrum $S\al{I}_{\rm min}(\epsilon)={\rm min}_{\theta_\pm,\xi_\pm}\ \av{\pq{ {J_{\xi_+,\xi_-}\al{\theta_+,\theta_-}}(\epsilon)}^2}$ is then equal to Eq.~\rp{Smin} evaluated for the parameters in Eq.~\rp{nnm1}, and it can be measured by minimizing the homodyne spectrum over the phases of both the local oscillator and of the filter. In particular, while the 
%minimization over 
tuning of the filter phases, $\varphi$ and $\varphi'$, can be, in principle, realized 
by recording the photocurrent for a sufficiently long time and then post processing the recorded signal,
%on a single experimental run, by the spectral analysis of the recorded time signal, 
the phases of the local oscillator, $\theta$ and $\theta'$, have to be adjusted during repeated homodyne measurements.
%minimization over $\theta_\pm$ is realized by repeating the homodyne measurement at different phases of the local oscillator.
%

\subsection{Two-mode homodyne spectrum and entangled spectral components}\label{2modehomodyne}

\begin{figure}[!t]
\begin{center}
\includegraphics[width=8.5cm]{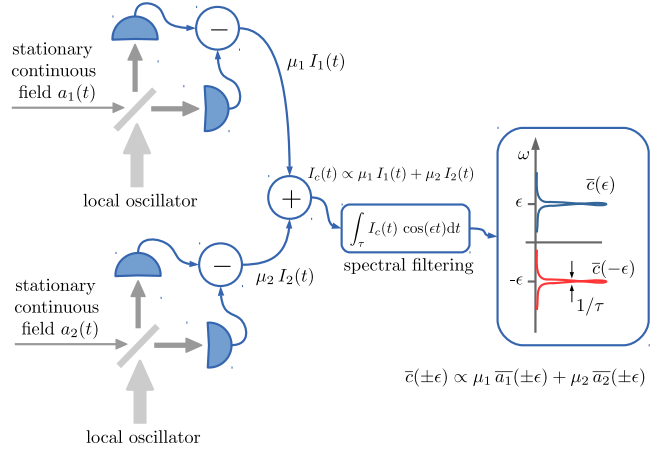}
\end{center}
\caption{Two-modes homodyne detection:
Two stationary continuous fields are detected by homodyne techniques. The two photocurrents are summed and then analyzed in frequency. As in the single-mode homodyne detection, the 
resulting total filtered photocurrent is a superposition of spectral modes  at opposite sideband frequencies $\pm\epsilon$. However, here, each spectral mode [$\overline c(\pm\epsilon)$] can be decomposed as the superposition of two spectral components [$\overline a_1(\pm\epsilon)$ and $\overline a_2(\pm\epsilon)$], at the same sideband frequency, each filtered from one of the two fields.
}\label{fig2}
\end{figure}

In the preceding section we have seen that the single-mode homodyne spectrum, which is measurable from a single stationary continuous field, provides informations about the entanglement between two spectral components. Let us now study the two-mode squeezing spectrum obtained from the combination of two homodyne photocurrents which result from the measurement of two signal fields $a_1(t)$ and $a_2(t)$, as depicted in Fig.~\ref{fig2}. This strategy detects the correlations between four spectral components~\cite{Schumaker}. 
Specifically, we consider the situation in which the photocurrents corresponding to the two detected fields, $I_{1}\al{\theta_1}(t)$ and $I_2\al{\theta_2}(t)$ have the form of Eq.~\rp{It0}, and are combined to construct the total photocurrent 
\begin{eqnarray}
I_c\al{\theta_1,\theta_2}(t)=\frac{1}{\sqrt{\mu_1^2+\mu_2^2}}\pq{\mu_1\,  {I_{1}\al{\theta_1}}(t)+\mu_2\,{I_{2}\al{\theta_2}}(t)} \ ,
\end{eqnarray}
where we have introduced the scaling parameters $\mu_1$ and $\mu_2$, which weight differently the two quadratures, and hence provide a means to select arbitrary collective modes as discussed below. 
These parameters are controllable experimentally including asymmetrical amplification and/or attenuation of the two photocurrents.
The total photocurrent is then analyzed in frequency.
Similarly to the previous case, the filtered photocurrent $ {J_c\al{\theta_1\theta_2}}(\epsilon)=\lim_{\tau\to\infty}
\frac{1}{\sqrt{\tau}}\int_{t-\tau}^t \dd s\   \cos(\epsilon\,s+\varphi)\ I_c\al{\theta_1\theta_2}(s)$ can be decomposed into two spectral components at the frequencies $\pm\epsilon$
\begin{eqnarray}\label{filterIc}
 {J_c\al{\theta_+,\theta_-}}(\epsilon)&=&\frac{1}{\sqrt{2}}\pq{ \overline{x_c\al{\theta_+}}(\epsilon) +\overline{x_c\al{\theta_-}}(-\epsilon)  }
\end{eqnarray}
where $\theta_\pm=\frac{\theta_1+\theta_2}{2}\pm\varphi$, and where we have introduced the quadrature operators for the collective spectral modes defined as
\begin{eqnarray}
\overline{x_c\al{\theta}}(\epsilon)=\ee^{\ii\theta}\overline{c}(\epsilon) 
+\ee^{-\ii\theta}{\overline{c\da}(-\epsilon)}
\end{eqnarray}
with the collective annihilation and creation operators given by
\begin{eqnarray}\label{barc} 
\overline{c}(\pm\epsilon)&=&\frac{1}{\sqrt{\mu_1^2+\mu_2^2}}\pq{\mu_1\,\ee^{\ii\theta_c}\,\overline {a_1}\pt{\pm\epsilon}+\mu_2\,\ee^{-\ii\theta_c}\,\overline {a_2}\pt{\pm\epsilon}} \ ,
\nn\\
\overline{c\da}(\mp\epsilon)&=&\frac{1}{\sqrt{\mu_1^2+\mu_1^2}}\pq{\mu_1\,\ee^{-\ii\theta_c}\,\overline {a_1\da}\pt{\mp\epsilon}+\mu_2\,\ee^{\ii\theta_c}\,\overline {a_2\da}\pt{\mp\epsilon}} \ .
\end{eqnarray}
where $\pq{\overline{c}(\pm\epsilon),\overline{c\da}(\mp\epsilon)}=1$ and
$\theta_c=\pt{\theta_1-\theta_2}/{2}$.
Also in this case, the filtered photocurrent is a composite quadrature of the form of Eq.~\rp{narrowI}. Thus, although it is constructed form the collective operators in Eq.~\rp{barc}, we can still apply the results of Sec.~\ref{squeezEnt} to conclude that  the two-mode squeezing spectrum, $S\al{II}(\epsilon)$, obtained as the minimum over the phases $\theta_\pm$ of the autocorrelation function of the filtered current, has the form of Eq.~\rp{S} but 
evaluated with the parameters
\begin{eqnarray}\label{nmc}
n_\pm\al{c}&=&\av{\overline{c\da}(\mp\epsilon)\, \overline{c}(\pm\epsilon)}
\nn\\&=&\frac{
\mu_1^2\, v\al{11}_\pm+\mu_2^2\,v\al{22}_\pm
+2\,\mu_1\,\mu_2\,\abs{v\al{21}_\pm}\,\cos\pq{2\,\theta_-+{\rm arg}(v\al{21}_\pm)}}
{\mu_1^2+\mu_2^2}
\nn\\
m\al{c}&=&\av{\overline{c}(\epsilon)\, \overline{c}(-\epsilon)}
\nn\\
&=&\frac{1}{\mu_1^2+\mu_2^2}\pq{\mu_1^2\,\ee^{\ii(\theta_1+\varphi)}w\al{11}+\mu_2^2\,\ee^{-\ii(\theta_1+\varphi)}w\al{22}
%}\nn\\&&\rpq{
+\mu_1\,\mu_2\pt{\ee^{\ii(\theta_2+\varphi)}w\al{12}+\ee^{-\ii(\theta_2+\varphi)}w\al{21}  }
}\ ,
\end{eqnarray}
where
 \begin{eqnarray}
v_\pm\al{jk}&=&\av{\overline{a_j\da}(\mp\epsilon)\,\overline{a_k}(\pm\epsilon)}
\nn\\
w\al{jk}&=&\av{\overline{a_j}(\epsilon)\,\overline{a_k}(-\epsilon)}
\end{eqnarray}
for $j,k= 1,2$. We remark that $S\al{II}(\epsilon)$ is found by minimizing the autocorrelation function of the photocurrent in Eq.~\rp{filterIc} over $\theta_\pm$ only, while $\theta_c$ and $\mu_j$ define the specific collective modes that are being probed and are fixed. 
Therefore, in this case the condition $S\al{II}(\epsilon)<1$ indicates the entanglement of the collective modes described by the operators in Eq.~\rp{barc}, each of which is a superpositions of the spectral components at the same sideband frequencies of the two fields. 
Moreover  similarly to the discussions of the single-mode homodyne spectrum, the corresponding logarithmic negativity can be, in turn, measured by summing two filtered photocurrent, at different phases, of the form of Eq.~\rp{filterIc},
and then calculating the corresponding autocorrelation function. The two-mode squeezing spectrum $S\al{II}_{\rm min}(\epsilon)$ is then found by minimizing this quantity over both the phases of the local oscillator and of the filter, and the result, in full similarity with the single-mode squeezing spectrum, is equal to Eq.~\rp{Smin}, but now evaluated for the parameters in Eq.~\rp{nmc}.

\subsection{Detecting single spectral components with homodyne techniques}
\label{single}

As discussed in the previous sections it is possible to construct arbitrary superposition of spectral modes at opposite  sideband frequencies by the superposition of two homodyne photocurrents. The corresponding total photocurrent is then given by Eq.~\rp{Ixi}. Similarly, when the phases  in Eq.~\rp{tetaxi} are set to some values for which one of the two parameters $\xi_\pm$ is equal to zero, then a single spectral mode is detected.

When applied to two distinct fields, this approach would permit the investigation of  the correlations between two distinct spectral modes each belonging to a different field.
Let us, for example, assume that we repeat the pair of measurements resulting in the total photocurrent in Eq.~\rp{Ixi} on two different continuous fields $a_1(t)$ and $a_2(t)$. 
The two resulting composite filtered photocurrents are then, in general, given by $ {J_{\xi_{+,j},\xi_{-,j}}\al{\theta_{+,j},\theta_{-,j}}}(\epsilon)
= 
\pq{\xi_{+,j} \overline{x\al{\theta_{+,j}}}(\epsilon)+ \xi_{-,j}\   \overline{x\al{\theta_{-,j}}}( -\epsilon)}/{\sqrt{\xi_{+,j}^2+\xi_{-,j}^2}}$ 
with the parameters defined as in Eq.~\rp{tetaxi}, and where, here, $j=1,2$ distinguish the parameters corresponding to the measurements of the first and of the second field respectively.

If, in each pair of measurements, we tune the phases of the local oscillators and of the filters
%If we now set the phases in each pair of measurement 
to certain values for which
${\theta_j-\theta_j'-(-1)^j(\varphi_j-\varphi_j')}=\pi$ and ${\theta_j-\theta_j'+(-1)^j(\varphi_j-\varphi_j')}\neq\pi$,  so that $\xi_{-,1}=\xi_{+,2}=0$, then each composite photocurrent is proportional to a quadrature of a single spectral component corresponding respectively to the annihilation operators
\begin{eqnarray}\label{aepsilon2}
\overline a_1(\epsilon)   \ \ \ \ \ \ {\rm and} \ \ \ \ \ \ \overline a_2(-\epsilon)\ .
\end{eqnarray} 
The two photocurrents are then summed together, after being multiplied by appropriately chosen scaling factors $\zeta_\pm$, so that the resulting total photocurrent is
\begin{eqnarray}
 {J_{tot}} 
&=& \frac{1}{\sqrt{\zeta_{+}^2+\zeta_{-}^2}}\pq{\zeta_{+} \overline{x_1\al{\theta_{+,1}}}(\epsilon)+\zeta_{-}\   \overline{x_2\al{\theta_{-,2}}}( -\epsilon)}\ ,
\nn
\end{eqnarray}
where the single mode quadratures $\overline{x_j\al{\theta}}(\pm\epsilon)$ are defined in Eq.~\rp{overlineX}.
Thus, this protocol detects the combined quadrature defined in Eq.~\rp{Zxi}. The corresponding squeezing  spectrum, $S\al{III}(\epsilon)$, 
defined for $\zeta_+=\zeta_-$, as the minimum of the autocorrelation function of the total photocurrent over $\theta_{+,1}$ and $\theta_{-,2}$ is equal to Eq.~\rp{S},
and is obtained by appropriately tuning the phases of the local oscillator and of the filter during repeated  homodyne measurements.
%, and is obtained by repeated homodyne measurements at different values of the phases of the local oscillator and of the filter.
Similarly, $S\al{III}_{\rm min}(\epsilon)$, defined as the minimum
of the power spectrum of the total photocurrent over $\theta_{+,1}$, $\theta_{-,2}$ and $\zeta_\pm$, is equal to Eq.~\rp{Smin}. In particular, also in this case we can conclude that this quantity is equivalent to the logarithmic negativity between $\overline a_1(\epsilon)$ and $\overline a_2(-\epsilon)$ when the fields are gaussian.

\subsection{Two-modes heterodyne spectrum and entangled spectral components}
\label{2modeheterodyne}

An alternative strategy to probe single spectral modes and hence to measure the squeezing spectrum, and the logarithmic negativity between two distinct spectral components of two distinct fields, as in Sec.~\ref{single}, is based on heterodyne measurements.

\begin{figure}[!t]
\begin{center}
\includegraphics[width=8cm]{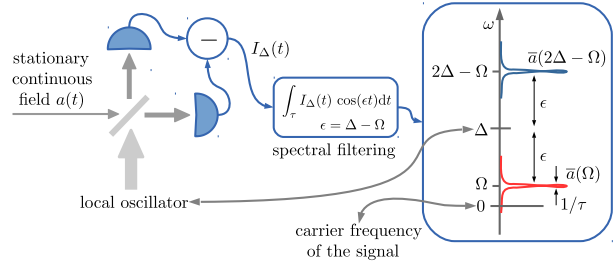}
\end{center}
\caption{Single-mode heterodyne detection: A stationary continuous field is detected by heterodyne techniques with the local oscillator at the frequency $\Delta$ relative to the carrier frequency of the signal. 
The detected photocurrent is spectrally analyzed at the frequency $\epsilon=\Delta-\Omega$. 
The resulting filtered photocurrent is a superposition of spectral components at the frequencies $\Omega$ and $2\Delta-\Omega$.
}\label{fig2b}
\end{figure}

In heterodyne techniques, the local oscillator is detuned from the carrier frequency of the signal field by a quantity $\Delta=\omega_{LO}-\omega_L$, and the corresponding operator for the photocurrent reads
$$I_\Delta\al{\theta}(t)=\ee^{\ii\Delta t}\ \ee^{\ii\theta}\ a(t) +\ee^{-\ii\Delta t}\ \ee^{-\ii\theta}\ a\da(t)\ .$$
Hence, the corresponding filtered photocurrent, $ {J_{\Delta}\al{\theta,\varphi}}(\epsilon)=\lim_{\tau\to\infty}\frac{1}{\sqrt{\tau}}\int_{t-\tau}^t\dd s\ \cos\pt{\epsilon\, s+\varphi}\ I_\Delta\al{\theta}(s)
$, is the superposition of the quadratures for the spectral components at the frequencies $\Delta\pm\epsilon$, namely 
$ {J_\Delta\al{\theta,\varphi}}(\epsilon)=\pq{ \overline{x\al{\theta+\varphi}}(\Delta+\epsilon)+ \overline{x\al{\theta-\varphi}}(\Delta-\epsilon)}/{\sqrt{2}}$.
Thus, while homodyne detection probes spectral components at opposite sideband frequencies, heterodyne techniques measure two components asymmetrically located with respect to the carrier frequency of the signal field, but symmetric with respect to the local oscillator frequency, that, in our description (where all frequencies are relative to the carrier signal), is equal to $\Delta$ (see Fig.~\ref{fig2b}).

In particular heterodyne techniques can be used to detect a single spectral component, as discussed in the following. Say we want to detect  the component at frequency $\Omega$, then we set the detuning $\Delta$ at a value much larger than the typical bandwidth of the signal $\abs{\Delta}\gg\Delta_{signal}$, that is the band of frequencies that are populated by the signal photons. We also assume that the frequency $\Omega$ is a relevant frequency for the field $\abs{\Omega}\leq \Delta_{signal}$. Then the corresponding heterodyne photocurrent is filtered at the frequency $\epsilon=\Delta-\Omega$, so that, as depicted in Fig.~\ref{fig2b}, ${J_\Delta\al{\theta,\varphi}}(\epsilon)$ is the superposition of the field quadratures at the frequencies $\Omega$ and $2\Delta-\Omega$
\begin{eqnarray}\label{narrowIhetero}
 {J_\Delta\al{\theta,\varphi}}(\Delta-\Omega)=\frac{1}{\sqrt{2}}\pq{ \overline{x\al{\theta-\varphi}}(\Omega)+ \overline{x\al{\theta+\varphi}}(2\,\Delta-\Omega)} \ .
\end{eqnarray}
Since the signal covers a bandwidth much smaller than $\Delta$, then the mode at $2\Delta-\Omega$ is basically in vacuum and only the photons of the sideband $\Omega$ are detected. However in doing this the vacuum fluctuations of the  empty component at $2\Delta-\Omega$ are added to the signal resulting in higher noise. 

This approach can  be exploited to measure the correlations between two spectral modes belonging to two separate fields. 
Specifically the two-modes heterodyne spectrum is obtained when detecting two signals with two heterodyne measurement (with $\Delta\gg\Delta_{signal}$).
The corresponding photocurrents are filtered independently at the frequencies $\epsilon_1$ and $\epsilon_2$ respectively, and then combined, with appropriately chosen scaling factors $\xi_j$,
in order to construct the total filtered photocurrent
$J_{\Delta,tot}\propto\xi_1{J_\Delta\al{\theta_1,\varphi_1}}(\epsilon_1)+\xi_2{J_\Delta\al{\theta_2,\varphi_2}}(\epsilon_2)$.
If, in particular, we are interested in the spectral components at frequency $\Omega$ of the first field, and at frequency $-\Omega$ of the second, such that $\abs{\Omega}\leq\Delta_{signal}$, we consider the filtered photocurrents at the frequencies
$\epsilon_1=\Delta-\Omega$ and $\epsilon_2=\Delta+\Omega$. Similarly to Eq.~\rp{narrowIhetero} they are equal, respectively, to the superpositions of the two quadratures at frequencies $\Omega$ and $2\,\Delta-\Omega$ of the first field, and of the two quadratures at frequencies $-\Omega$ and $2\,\Delta+\Omega$ of the second. Correspondingly, the total photocurrent is given by
\begin{eqnarray}
  {J_{\Delta,tot}}&=&
\frac{\xi_1\, \overline{x_1\al{\theta_1-\varphi_1}}(\Omega) +\xi_2\,\overline{x_2\al{\theta_2-\varphi_2}}(-\Omega) }{\sqrt{2}\sqrt{\xi_1^2+\xi_2^2}}
%\nn\\&&
+\frac{\xi_1\, \overline{x_1\al{\theta_1+\varphi_1}}(2\Delta-\Omega) +\xi_2\,\overline{x_2\al{\theta_2+\varphi_2}}(2\Delta+\Omega) }{\sqrt{2}\sqrt{\xi_1^2+\xi_2^2}}
\end{eqnarray}
where the quadrature operators for a single component are defined in Eq.~\rp{overlineX}. Its autocorrelation function is therefore given by
\begin{eqnarray}\label{GGheterodyne}
\av{\pq{  {J_{\Delta,tot}}}^2}
&=&\frac{1}{2}\av{\pq{X\al{\theta_{1}-\varphi_1,\theta_{2}-\varphi_2}_{\xi_1,\xi_2}(\Omega)}^2}+\frac{1}{2}
\nn\\
\end{eqnarray}
where the term $\frac{1}{2}$ in the right hand side is due to the vacuum fluctuations of the spectral modes at $2\Delta\pm\Omega$, the collective quadrature $X\al{\theta_{1},\theta_2}_{\xi_{1},\xi_2}(\Omega)$ has the same form of the one  defined in Eq.~\rp{Zxi}, and its autocorrelation function, that is given in Eq.~\rp{DeltaXxi}, quantifies the correlations between the modes $\overline{a_1}(\Omega)$ and $\overline{a_2}(-\Omega)$.

Also in this case we define two kinds of optimized squeezing spectra.
One is obtained by minimizing the autocorrelation function in Eq.~\rp{GGheterodyne}, with $\xi_1=\xi_2$, only over the phases of the local oscillators $T(\Omega)={\rm min}_{\theta_j}\ \av{\pq{  {J_{\Delta,tot}}}^2}$; the other is obtined when the minimization runs also over the scaling parameters, $T_{\rm min}(\Omega)={\rm min}_{\theta_j,\xi_j}\ \av{\pq{ {J_{\Delta,tot}}}^2}$.
In both cases they can be expressed in terms of the squeezing spectra resulting form the protocol described in Sec.~\ref{single}, as
\begin{eqnarray}
T(\Omega)=\frac{S\al{III}(\Omega)+1}{2}\ ,
\ \ \ \ \ \ \ \ \ 
T_{\rm min}(\Omega)=\frac{S_{\rm min}\al{III}(\Omega)+1}{2}\ .
\end{eqnarray}
Thereby, according to Eq.~\rp{nuOmega}, in the gaussian case, $T_{\rm min}(\Omega)$ measures the entanglement between the modes whose operators are $\overline a_1(\Omega)$ and $\overline a_2(-\Omega)$.

\section{Application to ponderomotive squeezing in a two sided cavity}\label{ponderomotive}

Here we study ponderomotive squeezing~\cite{Fabre,Mancini94,Brooks,Safavi-Naeini} and the conditions under which the spectral components of the field emitted by an optomechanical system are squeezed and entangled.  Moreover we determine the squeezing spectra, and we identify the detectable spectral modes that exhibit maximum entanglement.

\begin{figure*}[!t]
\begin{center}
\includegraphics[width=18cm]{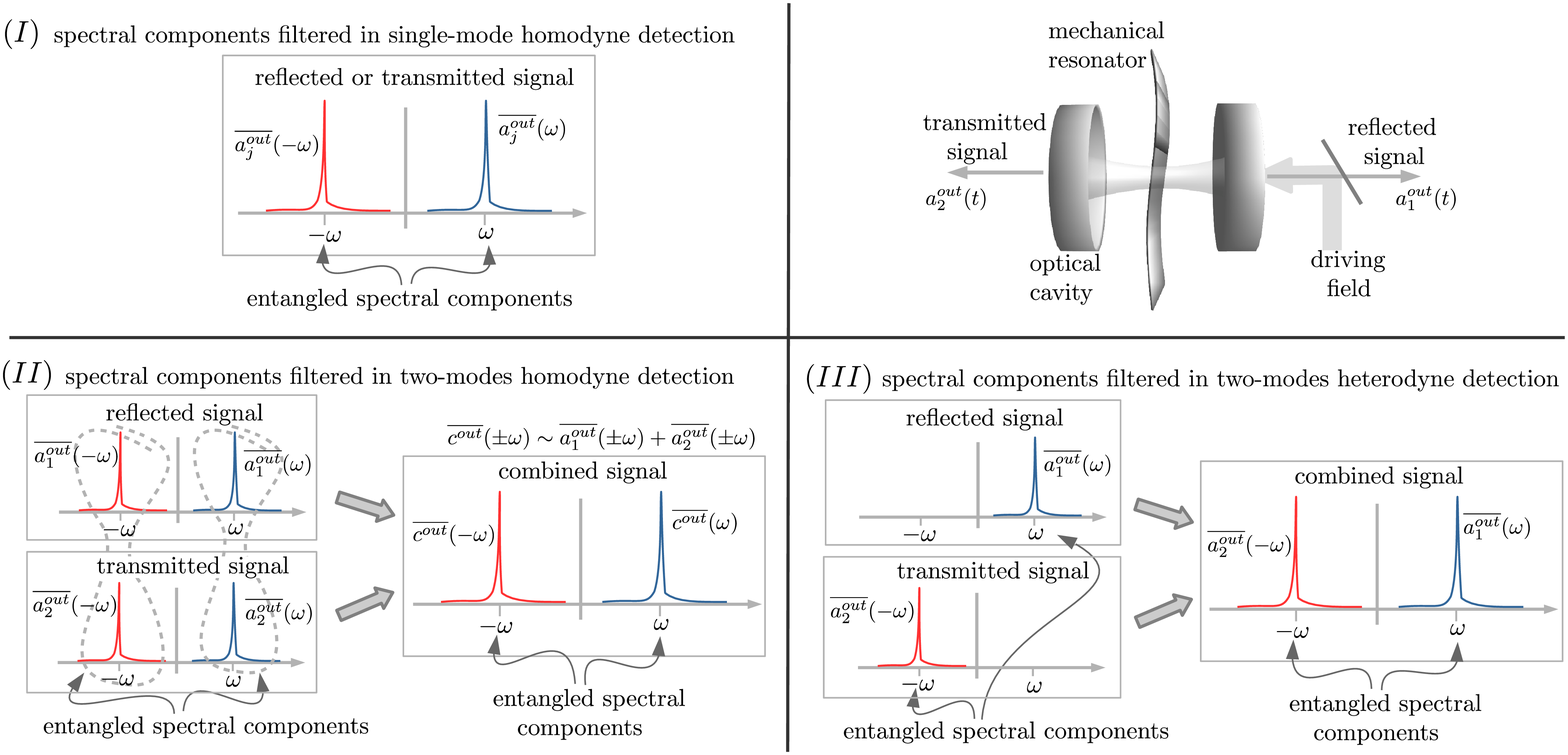}
\end{center}
\caption{
Spectral components at the output of an optomechanical system probed with three different detection strategies and that are entangled, and squeezed, as a result of the optomechanical interaction. 
}\label{fig3}
\end{figure*}

Ponderomotive squeezing refers to the squeezing of the output light resulting from the non-linear radiation-pressure interaction with a mechanical resonator inside an optical cavity. 
The response of a high-Q mechanical resonator to a resonance mode of a high-finesse optical cavity can be described as that of a Kerr medium, which imparts an intensity-dependent phase shift to the light. 
As a result the field fluctuations can be reduced and, correspondingly, squeezed light is produced~\cite{Fabre,Mancini94}.
In detail, we investigate a Fabry-Perot cavity with a membrane in the middle~\cite{Biancofiore}. 
A single optical mode is relevant in the system dynamics. It loses photons at rates $\kappa_j$ from the mirrors $j=1,2$, and is driven by a laser at a frequency detuned by $\delta$ from the relevant cavity resonance. Only one mechanical mode of the membrane at frequency $\omega_m$ interacts significantly with the cavity field  with a linearized coupling strength $g$. 
The decay rate of the membrane  is $\gamma$, and the number of thermal mechanical excitations $n_T$.
The corresponding linearized optomechanical dynamics is Gaussian~\cite{Genes,Aspelmeyer}  and is efficiently analyzed in terms of the standard input-output theory~\cite{Tufarelli}. 
Here we describe the results for the field emitted through the two cavity mirrors at the steady state of the system dynamics in the regime of optomechanical stability, referring  the reader to the App.~\ref{inputoutput} for further details and derivations.

The photons lost through the two cavity mirrors are described by the output field operators $a^{out}_1$, ${a^{out}_1}\da$, $a^{out}_2$, and ${a^{out}_2}\da$. The corresponding  power spectrum matrix, defined in Eq.~\rp{Aom}, can be evaluated for the vector of operators  $\va_{out}=\pt{a^{out}_1,a^{out}_2,{a^{out}_1}\da,{a^{out}_2}\da}^T$, and 
the result is given by 
\begin{eqnarray}\label{AAAout}
\widetilde\PP_{out}(\omega)
&=&\frac{2\, g^2}{\abs{f(\omega)}^2}\ \QQ_{out}\ \WW\ \QQ_{out} +\YY 
\end{eqnarray}
where $\QQ_{out}$ is a diagonal matrix whose diagonal elements are 
$\pt{\sqrt{2\kappa_1},\sqrt{2\kappa_2},\sqrt{2\kappa_1},\sqrt{2\kappa_2}}$,
$\YY$ is a matrix whose only non zero elements are $\pg{\YY}_{1,3}=\pg{\YY}_{3,4}=1$,
\begin{eqnarray}
f(\omega)&=& 4\, g^2\,\delta\,{\omega_m}-\pq{{\omega_m}^2+\pt{\gamma-\ii\omega}^2}\pq{\delta^2+\pt{\kappa_1+\kappa_2-\ii\omega}^2}\ ,
\nn\\
\end{eqnarray}
and the matrix $\WW$ is given in terms of the parameters
\begin{eqnarray}
\alpha&=&
-4\,g^2\,{\omega_m}^2\pt{\kappa_1+\kappa_2+\ii\delta}-\pq{\pt{\kappa_1+\kappa_2+\ii\delta}^2+\omega^2}
%\nn\\&&\times
\pq{\gamma\pt{2n_T+1}\pt{\gamma^2+{\omega_m}^2+\omega^2} +\ii{\omega_m}\pt{\gamma^2+{\omega_m}^2-\omega^2} }
\nn\\
\beta_{\pm\omega}&=&
4\,g^2\,{\omega_m}^2\,\pt{\kappa_1+\kappa_2}+\gamma\pq{(\kappa_1+\kappa_2)^2+(\delta\pm\omega)^2}
%\nn\\&&\times
\pq{\pt{2n_T+1}\pt{\gamma^2+{\omega_m}^2+\omega^2}\mp2\,\omega\,{\omega_m}}
\nn\\
\end{eqnarray}
(with $\alpha$ complex even function of $\omega$, and $\beta_{\pm\omega}$ real and positive) as
\begin{eqnarray}
\WW=\pt{
\begin{array}{cccc}
\alpha^*&\alpha^*&\beta_{\omega}&\beta_{\omega}\\
\alpha^*&\alpha^*&\beta_{\omega}&\beta_{\omega}\\
\beta_{-\omega}&\beta_{-\omega}&\alpha&\alpha\\
\beta_{-\omega}&\beta_{-\omega}&\alpha&\alpha
\end{array}
}\ .
\end{eqnarray}
This matrix contains all the informations about the spectral properties of the output fields and can be used to construct the correlation matrix for two spectral modes as in Eq.~\rp{AOO}. In the following we will use this matrix and the results of sections~\ref{squeezEnt} and \ref{homodyne-heterodyne} to study the corresponding entanglement properties.

\begin{figure*}[!th]
\begin{center}
\includegraphics[width=17cm]{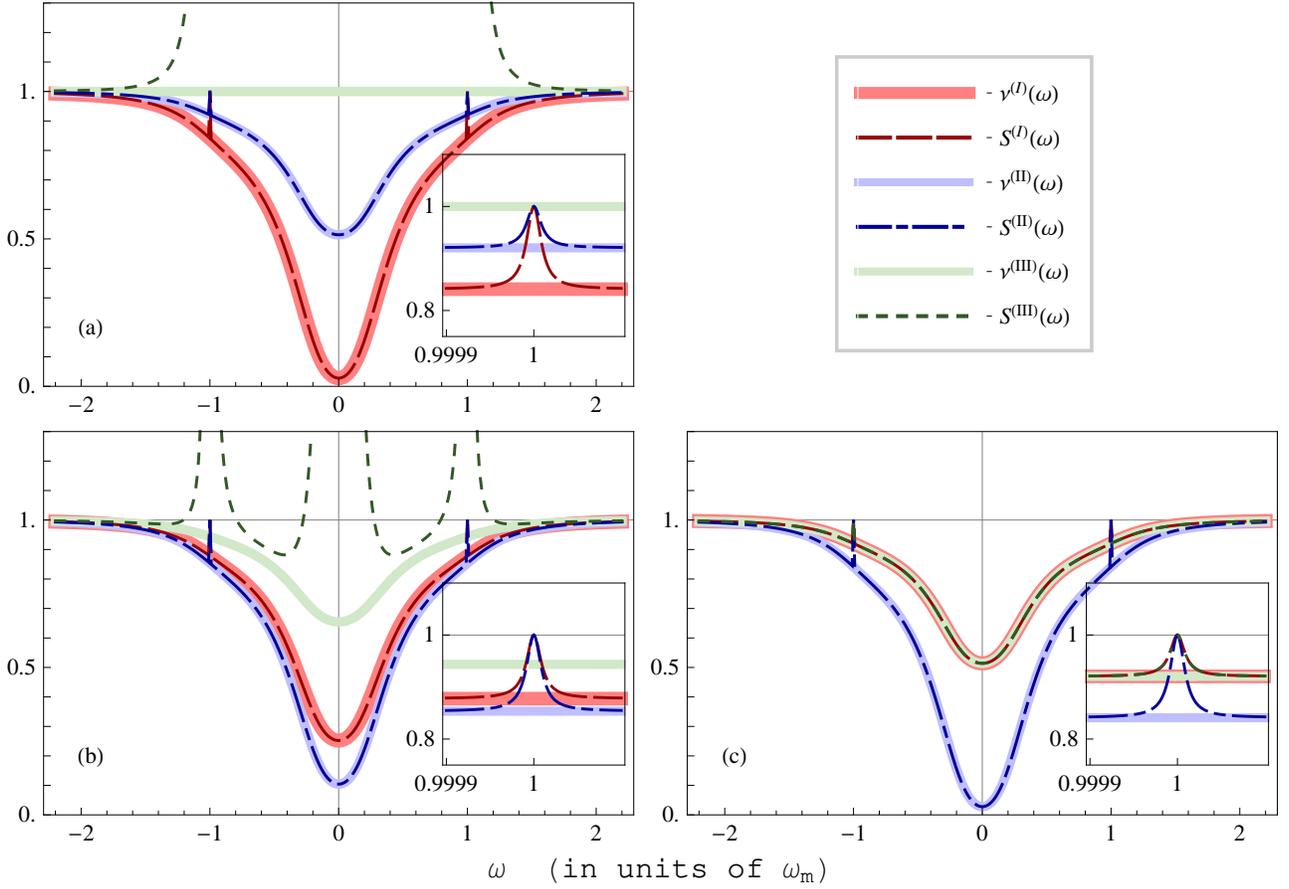}
\end{center}
\caption{
Squeezing spectrum $S\al{\ell}(\omega)$ and symplectic eigenvalue $\nu\al{\ell}(\omega)=S\al{\ell}_{\rm min}(\omega)$,
obtained for the values of the decay rates of the cavity mirrors $\kappa_1+\kappa_2=0.1\omega_m$ and (a) $\kappa_2=0$, (b) $\kappa_2/\kappa_1=0.3$, (c) $\kappa_2/\kappa_1=1$.
In the insets the regions close to the upper mechanical resonance ($\omega=\omega_m$) are magnified.
The other parameters are 
$\delta=0$, $g=0.5\,\omega_m$, $\gamma=10^{-5}\,\omega_m$, $n_T=13091$ (temperature$\,=100\,$mK and $\omega_m=1\,$MHz), and in the case of $S\al{II}(\omega)$ and  $\nu_-\al{II}(\omega)$, the spectral mode operators $\overline c^{out}(\pm\epsilon)$ are defined by the values $\theta_c=0$ and $\mu_2=\mu_1$.
}\label{figS2}
\end{figure*}
\begin{figure}[!th]
\begin{center}
\includegraphics[width=7.5cm]{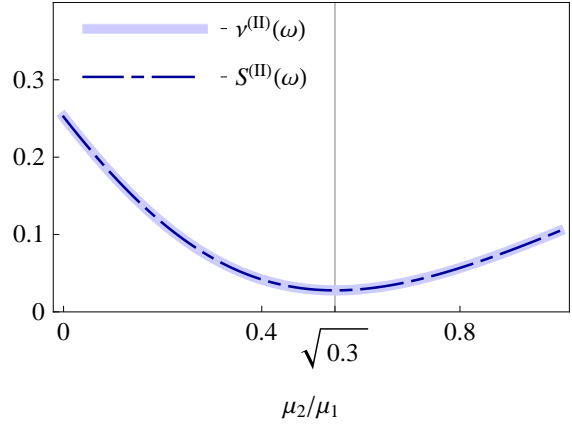}
\end{center}
\caption{
Squeezing spectrum $S\al{II}(0)$ and symplectic eigenvalue $\nu\al{II}(0)=S\al{II}_{\rm min}(0)$, as a function of the ratio $\mu_2/\mu_1$, when $\theta_c=0$, $\kappa_2/\kappa_1=0.3$, $\kappa_1+\kappa_2=0.1\omega_m$, 
$\delta=0$, $g=0.5\,\omega_m$, $\gamma=10^{-5}\,\omega_m$, $n_T=13091$ (temperature$\,=100\,$mK and $\omega_m=1\,$MHz).
}\label{figS3}
\end{figure}

\subsection{Homodyne and heterodyne spectra and entangled components of the emitted field}

In Sec.~\ref{homodyne-heterodyne} we have described three different strategies for the experimental investigation of the spectral properties of stationary continuous fields, which are based on homodyne and heterodyne techinques. They probe different  pairs of spectral modes given, respectively, by Eqs.~\rp{aepsilon1}, \rp{barc} and \rp{aepsilon2}. When applied to the investigation of the optomechanical system, these techniques allow the detection of the corresponding spectral components of the two output fields $a_j^{out}(t)$ as depicted in Fig.~\ref{fig3}. In particular, using these techniques it is possible to study composite quadratures of these pairs of modes and their squeezing and entanglement properties. We have identified two different kinds of optimized squeezing spectra, corresponding to different experimental approaches for the measurement and the minimization of the homodyne photocurrent fluctuations. Specifically, in order to probe symmetric superpositions of quadratures of  the two modes it is sufficient to apply standard homodyne techniques, thereby the minimization is realized tuning the relative phase of the two quadratures, which is controlled experimentally by the phase of the local oscillator. We indicate this phase-optimized spectrum with the symbol $S\al{\ell}(\omega)$ where the label $\ell=I,II,III$ is used to distinguish the three detection strategies (see Fig.~\ref{fig3}). 
On the other hand,  non-symmetric superpositions, with different weights of the two quadratures, can be probed by combining different filtered photocurrent detected with appropriately selected phases of both the filter and the local oscillator. The globally-optimized squeezing spectrum $S\al{\ell}_{\rm min}(\omega)$ is then obtained minimizing the corresponding fluctuations over both the phases of the local oscillator and of the filter.

In all cases the squeezing spectra $S\al{\ell}(\omega)$ and  $S\al{\ell}_{\rm min}(\omega)$ are equal to Eq.~\rp{S} and \rp{Smin}, evaluated in each case for the specific parameters $n_\pm$ and $m$ which correspond to the detected spectral modes.
Specifically, $S\al{I}(\omega)$ and  $S\al{I}_{\rm min}(\omega)$ are obtained by single-mode homodyne detection of a single field (either one of the two output fields), as discussed in Sec.~\ref{singelhomodyne}, and if applied to the output from the first mirror then $n\al{I}_\pm=\av{\overline{{a_1^{out}}\da}(\mp\omega)\ \overline{a_1^{out}}(\pm\omega) }$ and $m\al{I}=\av{\overline{{a_1^{out}}}(\omega)\ \overline{a_1^{out}}(-\omega) }$. The second strategy is based on the two-mode homodyne detection of the two output fields (see Sec.~\ref{2modehomodyne}) and the corresponding spectra, $S\al{II}(\omega)$ and  $S\al{II}_{\rm min}(\omega)$, are evaluated for $n\al{II}_\pm=\av{\overline{{c_j^{out}}\da}(\mp\omega)\ \overline{c_j^{out}}(\pm\omega) }$ and $m\al{II}=\av{\overline{{c_j^{out}}}(\omega)\ \overline{c_j^{out}}(-\omega) }$, where the operators $\overline{c\al{out}}(\pm\omega)$ have the same form of Eq.~\rp{barc} but, in this case, they are constructed as the superpositions of the two filtered output fields $\overline{a_j^{out}}(\pm\omega)$. Finally,  $S\al{III}(\omega)$ and  $S\al{III}_{\rm min}(\omega)$, correspond to the two-modes heterodyne detection of the two output fields as discussed in Sec.~\ref{2modeheterodyne}. If it is applied to the spectral component at frequency $\omega$ of the first output and at $-\omega$ of the second then,  $S\al{III}(\omega)$ and  $S\al{III}_{\rm min}(\omega)$ are evaluated for $n\al{III}_\pm=\av{\overline{{a_1^{out}}\da}(\mp\omega)\ \overline{a_2^{out}}(\pm\omega) }$ and $m\al{III}=\av{\overline{{a_1^{out}}}(\omega)\ \overline{a_2^{out}}(-\omega) }$. We remark that these last two spectra can also be retrieved by combining various homodyne photocurrents as discussed in Sec.~\ref{single}.

The power spectrum matrix in Eq.~\rp{AAAout} can be used to find
\begin{eqnarray}
S\al{\ell}(\omega)&=&1+\frac{4\, g^2}{\abs{f(\omega)}^2}
\pq{
\abs{q_+\al{\ell}}^2\,\beta_\omega+\abs{q_-\al{\ell}}^2\,\beta_{-\omega}-2
\abs{q_+\al{\ell}\,q_-\al{\ell}\,\alpha}
} \ ,\nn\\
S_{\rm min}\al{\ell}(\omega)&=&1+\frac{4\, g^2}{\abs{f(\omega)}^2}
\pq{
\abs{q_+\al{\ell}}^2\,\beta_\omega+\abs{q_-\al{\ell}}^2\,\beta_{-\omega}
%}\nn\\&&\rpq{
-\sqrt{4\abs{q_+\al{\ell}\,q_-\al{\ell}\,\alpha}^2+\pt{\abs{q_+\al{\ell}}^2\,\beta_\omega-\abs{q_-\al{\ell}}^2\,\beta_{-\omega}}^2}
} \nn
\end{eqnarray}
where
\begin{eqnarray}
&&q_+\al{I}=q_-\al{I}=\sqrt{\kappa_1}
\nn\\
&&q_+\al{II}=q_-\al{II}=\frac{\mu_1\,\ee^{\ii\theta_c}\sqrt{\kappa_1}+\mu_2\,\ee^{-\ii\theta_c}\sqrt{\kappa_2}}{\sqrt{\mu_1^2+\mu_2^2}}
\nn\\
&&q_+\al{III}=\sqrt{\kappa_1}\ , \ \ \ \ \ \ \ \ \ \ q_-\al{III}=\sqrt{\kappa_2}\ ,
\end{eqnarray}
In the case of the strategy $(II)$, the parameters $\mu_j$ and $\theta_c$, in the expression for $q_\pm\al{II}$, determine the specific detected composite modes, that are defined as in Eq.~\rp{barc}.
We observe that when  
\begin{eqnarray}\label{maxII}
\theta_c=0  \ \ \ \ \ \ {\rm and}  \ \ \ \  \ \ \ \mu_1/\mu_2=\sqrt{\kappa_1/\kappa_2}
\end{eqnarray}
then $q_\pm\al{II}=\kappa_1+\kappa_2$, hence we can conclude that $S\al{II}_{\rm min}(\omega)$
evaluated for a two-sided configuration with decay rates $\kappa_1$ and $\kappa_2$, is equal to
$S\al{I}_{\rm min}(\omega)$
when it is evaluated 
for a single-sided cavity with the decay rate equal to the sum of the decay rates $\kappa_1+\kappa_2$ of the two-sided configuration.
It indicates that the entanglement between $\overline{a_1^{out}}(\omega)$ and $\overline{a_1^{out}}(-\omega)$, in a single-sided cavity, is redistributed between the four spectral components  $\overline{a_1^{out}}(\pm\omega)$ and $\overline{a_2^{out}}(\pm\omega)$ in the case of a two-sided cavity. For this reason, the same amount of squeezing found in the case of a single-sided cavity, can be recovered when the informations from the two decay channels of a two-sided cavity are properly combined. In particular $q_\pm\al{II}$ is maximum for the parameters of Eq.~\rp{maxII}, and consequently the corresponding modes are the collective modes that are maximally squeezed (and entangled).

We also note that according to Eq.~\rp{Ennm} we find that,  in all cases, the pairs of spectral components are entangled (and squeezed), although possibly with different degree of entanglement (and squeezing), when 
\begin{eqnarray}
{ \beta_{\omega}\,\beta_{-\omega}}<{\abs{\alpha}^2} \ .
\end{eqnarray}
Moreover we remark that the logarithmic negativity between the pair of spectral components detected with each strategy is obtained applying the definition in Eq.~\rp{logneg0} to the parameter $$\nu\al{\ell}(\omega)=S\al{\ell}_{\rm min}(\omega)\ ,$$ which is equal to the minimum symplectic eigenvalue of the corresponding partially transposed covariance matrix.

In Fig.~\ref{figS2} we compare the results for the spectra evaluated for realistic parameters and corresponding to the three detection strategies, and hence to different pair of spectral components. Each plot in Fig.~\ref{figS2} is evaluated for different values of the relative decay rate $\kappa_2/\kappa_1$  of the two mirrors, while the total decay rate $\kappa_1+\kappa_2$ and all the other parameters are kept fixed. In plot (a) we study a single sided cavity with $\kappa_2=0$. In plot (b) the mirrors are lossy and non-symmetric, and finally in (c) the two mirrors are symmetric $\kappa_1=\kappa_2$. When $\kappa_2=0$, in plot (a), the curves for $S\al{II}$, $\nu\al{II}$, $S\al{III}$ and $\nu\al{III}$ correspond to the situation in  which the detector on the second mirror detects only vacuum fluctuations. It is therefore clear that the curves for $S\al{III}$ and $\nu\al{III}$ that measures the correlations between the single spectral component $\overline{a_1^{out}}(\omega)$ of the field lost form the first mirror and the single spectral component $\overline{a_2^{out}}(-\omega)$ of the field lost from the second show no squeezing. On the other hand maximum two-mode squeezing and entanglement, is observed for the single-mode homodyne spectra corresponding to the curves  $S\al{I}$ and $\nu\al{I}$ that measure the correlations between $\overline{a_1^{out}}(\omega)$ and $\overline{a_1^{out}}(-\omega)$. The curves $S\al{II}$, $\nu\al{II}$, that measure the correlations between the composite modes in Eq.~\rp{barc}, are at an intermediate value as a result of the vacuum fluctuations of the modes $\overline{a_2^{out}}(\pm\omega)$ which reduces the visibility of the two-mode squeezing between  $\overline{a_1^{out}}(\omega)$ and $\overline{a_1^{out}}(-\omega)$.

Plot (c) corresponds to a symmetric two-sided cavity with $\kappa_1=\kappa_2$. In this case the maximum squeezing and entanglement is obtained for the curves $S\al{II}$ and $\nu\al{II}$, indicating that in this configuration the maximally entangled spectral components correspond to the composite modes in Eq.~\rp{barc}. In particular $S\al{II}$ and $\nu\al{II}$ are equal to the curves for $S\al{I}$ and $\nu\al{I}$ in the single-sided cavity reported in plot (a). In the two cases, the correlations of the cavity field built by the optomechanical dynamics are equal, as a result of the equal total decay rate. The only difference is that, in one case all the photons are lost through a single mirror, while in the other the they are split in the two decay channels and only their combined detection can reveals the corresponding total degree of squeezing and entanglement. Moreover in plot (c) we observe that $S\al{III}$ and $\nu\al{III}$ are equal to $S\al{I}$ and $\nu\al{I}$. In this case, the two output fields are symmetric, hence the correlations between $\overline{a_1\al{out}}(\omega)$ and $\overline{a_1\al{out}}(-\omega)$ are equal to that between $\overline{a_1\al{out}}(\omega)$ and $\overline{a_2\al{out}}(-\omega)$. Finally, plot (b) corresponds to an intermediate situation between the two described in (a) and (c), and shows that the three detection strategies can display different degrees of squeezing.  In general, the values of all the squeezing spectra can lay at any value between the extremes, set by the corresponding curves in plot (a) and (c), depending on the actual value of the ratio $\kappa_2/\kappa_1\in\pq{0,1}$. 

We also emphasize that the values of $S\al{II}$ and $\nu\al{II}$ are reported, in the three plots, for the same values of the parameters $\mu_j$ and $\theta_c$ which define the specific superposition of spectral components that are being probed as defined  in Eq.~\rp{barc}. However for each value of $\kappa_2/\kappa_1$ the values of $\mu_j$ and $\theta_c$ can be appropriately tuned in order to find the composite modes that are characterized by the same maximum amount of squeezing in as in plot (c). This is shown for the parameters of plot (b) and at $\omega=0$ in Fig.~\ref{figS3} where maximum of squeezing and entanglement (recovering the maximum value found in Fig.~\ref{figS2} (c)) is found when $\mu_2/\mu_1=\sqrt{\kappa_2/\kappa_1}$. We further observe that, in Fig.~\ref{figS2}, the results for $S\al{I}$ and  $S\al{II}$ are very close to $\nu\al{I}$ and $\nu\al{II}$ respectively. They are sensibly different only close to the mechanical resonances ($\omega=\pm\omega_m$), where although the two spectral modes are entangled ($\nu\al{\ell}<1$), this feature is not reflected in the corresponding squeezing spectrum $S\al{\ell}$, which is sensibly larger and very close to one. This happens when the two corresponding spectral modes are significantly asymmetric so that $n_+\al{\ell}\neq n_-\al{\ell}$. This situation is realized very close to the condition $\omega=\pm\omega_m$ and for a bandwidth of the order of the mechanical dissipation rate $\gamma$ which, in typical optomecanical system, can be very small as shown in the insets of Fig.~\ref{figS2}.  On the other hand, the discrepancy between $S\al{III}$ and $\nu\al{III}$ can be considerably larger, covering, for example, the full spectrum in plots (a) and (b). This is due to the fact that, in this case the difference between the corresponding $n_+\al{III}$ and $n_-\al{III}$ is proportional to the difference between $\kappa_1$ and $\kappa_2$, which is relatively large in plots (a) and (b). In plot (c), on the contrary, the two mirrors are symmetric and correspondingly $S\al{III}$ and $\nu\al{III}$ are very close, reproducing the results for $S\al{I}$ and $\nu\al{I}$.

\section{Conclusions}
\label{Conclusions}

In conclusion, we have presented a comprehensive analysis of the entanglement properties of stationary squeezed fields at the output of a quantum optical system. 
By revisiting a number of already known concepts and condensing them into a unified description, we have derived novel results that directly link the spectral properties of squeezed light fields, in the stationary continuous-wave regime, to the entanglement theory of continuous-variable systems.
Specifically we have employed long-time filtered modes to systematically study the spectral properties of squeezed fields. Correspondingly we have derived general squeezing and entanglement criteria valid for stationary fields and most importantly, we have established the equivalence between two-mode squeezing variance and logarithmic negativity for stationary Gaussian fields. 
In experiments, the squeezing properties of the field can be investigated with  homodyne or heterodyne techniques. In particular the long time integration of the homodyne or heterodyne signal, provides informations about specific spectral components of the field and of the corresponding squeezing. We have analyzed the discrete bosonic operators describing such spectral modes and we have studied the corresponding entanglement properties, thereby demonstrating that the measurable squeezing spectrum resulting from the spectral homodyne or heterodyne analysis of the field is, indeed a direct measurement of the corresponding logarithmic negativity.

When applied to an optomechanical system comprising a two-sided Fabriy-Perot cavity with a membrane in the middle,  these findings help in identifying the specific spectral components of the output fields that are maximally entangled, showing, in particular, that maximum squeezing and entanglement is found between specific modes constituted by the superposition of carefully selected  spectral components of the two outputs. 

In general, a continuous-wave squeezed field combines, in a single spatial mode, a large number of spectral entangled sideband modes. 
It is, therefore, logical to ask, whether and how one could exploit such rich entanglement structure for real quantum-enhanced applications. Such question has been addressed, for example in  Ref.~\cite{Zhang,Huntington,Glockl,Hage}, where it is discussed how to spatially separate the spectral sidebands of a continuous squeezed field in order to create $N$ spatially independent entangled pairs and, hence, to prepare $N$ quantum communication channels, whose actual number is limited only by the spectral resolution of the experimental apparatus and by the bandwidth of the squeezed signal. On a similar perspective, it is intriguing to ask if such large amount of  entangled pairs, could be exploited
as a resource for frequency encoded multimode entangled networks, in the stationary continuous-wave regime, alternative to that realized with pulsed frequency combs~\cite{Roslund}.  

We finally remark that although, here we have focused on spectral components, the approach based on the filtered modes that we have described in Sec.~\ref{filteresmodes} is sufficiently general to be 
applicable to a wider area of experimental situations in which finite-time filtered mode are relevant~\cite{Yurke87,Opatrny,Sasaki,Wasilewski06,Vitali08}.

\section*{Acknowledgments}
This work has been supported by the European Commission (ITN-Marie Curie project cQOM and FET-Open Project iQUOEMS) and by MIUR (PRIN 2011).

\appendix

\section{Squeezing and entanglement of discrete modes}\label{discrete}

Here we discuss some useful results regarding the squeezing and the entanglement of discrete Gaussian modes~\cite{Braunstein,Adesso07,Weedbrook}.  Squeezing refers to the occurrence of reduced fluctuations of a quadrature of the field below the value of the fluctuations of a coherent state. Particularly interesting is two-mode squeezing that refers to the squeezing of a combined quadrature of two modes, while the two separated modes are not squeezed. Two-mode squeezing is, in fact, a signature of entanglement between the two modes.

Let us now consider two discrete modes with annihilation operators, $b_1$ and $b_2$, for which $\pq{b_j,b_k\da}=\delta_{j,k}$. Here and in the following, for simplicity, we assume that the average value of the fields is zero $\av{b_j}=0$, and only the fluctuations characterize the state of the two modes. A quadrature $x_j\al{\phi}=\ee^{\ii\phi}b_j+\ee^{-\ii\phi}b_j\da$ is squeezed when the following relation is fulfilled
$\Delta x_j\al{\phi}=\av{\pq{x_j\al{\phi}}^2}<1$. Two mode squeezing is, similarly, found when 
the variance of a composite quadrature $X\al{\phi_1,\phi_2}=\frac{1}{\sqrt{2}}\pq{\ee^{\ii\phi_1}\,b_1+\ee^{-\ii\phi_1}\,b\da_1+\ee^{\ii\phi_2}\,b_2+\ee^{-\ii\phi_2}\,b\da_2}$ is smaller than one, i.e. $\Delta X\al{\phi_1,\phi_2}=\av{\pq{X\al{\phi_1,\phi_2}}^2
}<1$. We note that this relation can be satisfied only if $\av{b_1\ b_2}\neq 0$.
In  general, Two-mode squeezing variances of this form can be used to construct entanglement
criteria. In particular, it was established~\cite{Duan} that, given a quadrature of the form 
\begin{eqnarray}\label{Xcomp}
X_{\xi_1,\xi_2}\al{\phi_1,\phi_2}=
\frac{\xi_1\ee^{\ii\phi_1}\,b_1+\xi_1\ee^{-\ii\phi_1}\,b\da_1+\xi_2\ee^{\ii\phi_2}\,b_2+\xi_2\ee^{-\ii\phi_2}\,b\da_2}{\sqrt{\xi_1^2+\xi_2^2}}
\nn\\
\end{eqnarray}
where $\xi_j$ are real and positive.
A sufficient condition for entanglement can be defined in terms of the quantity
\begin{eqnarray}\label{entS}
E_S=\Delta X\al{\phi_1,\phi_2}_{\xi_1,\xi_2}+
\Delta X\al{\phi_1+\frac{\pi}{2},\phi_2-\frac{\pi}{2}}_{\xi_1,\xi_2} \ .
\end{eqnarray}
Specifically, when $E_S<2$, for some values of $\xi_j$, and $\phi_j$, then the two modes are entangled. In the case of Gaussian fields this criterion becomes also a necessary condition for entanglement (for appropriate values of $\xi_j$)~\cite{Braunstein}.

In the analysis of the entanglement properties of Gaussian systems, for which all the informations are contained in the first and second moments of the field operators, it is useful to introduce the following matrix notation. We consider the column vector of operators $\vb=(b_1,b_2,b_1\da,b_2\da)^T$ and the corresponding correlation matrix which is given by $$\AAA=\av{\vb\ \vb^T},$$ whose elements are $\pg{\AAA}_{j,k}=\av{\pg{\vb}_j\ \pg{\vb}_k}$. 
The corresponding covariance matrix, $C$, namely the symmetric matrix of correlations of the quadrature $x_j\al{0}$ and $x_j\al{\pi/2}$, can be used to compute entanglement measures, such as the logarithmic negativity. It is given by $C=\TT\frac{\AAA+\AAA^T}{2}\TT^T$, where we have introduced the matrix 
$$\TT=\pt{\small\mat{cccc}{ 1 &  0 & 1 & 0  \\
 -\ii &  0 & \ii & 0  \\
 0 &  1 & 0 & 1  \\
 0 &  -\ii & 0 & \ii   }} \ .$$
The logarithmic negativity for Gaussian states is then computed as $E_N={\rm max}\pg{0,-\log_2(\nu)}$, where $\nu$ is the smallest symplectic eigenvalue of the covariance matrix of the partially transposed state, that can be expressed as $\CC'=\Pi\,\CC\,\Pi$, where $\Pi$ is a diagonal matrix whose diagonal elements are $\pt{1,-1,1,1}$~\cite{Vidal02}. In particular, this relation implies that the state is entangled when $\nu<1$.

A generic covariance matrix can always be transformed, using only local symplectic transformation, into the standard form~\cite{Duan}
\begin{eqnarray}\label{standardC}
\CC_0=\pt{\mat{cccc}{
a&0&c&0\\
0&a&0&c'\\
c&0&b&0\\
0&c'&0&b
}}\ ,
\end{eqnarray}
where $a$, $b$, $c$ and $c'$ are reals.
In this case the corresponding matrix of correlations for the field operators reads
\begin{eqnarray}\label{standardA}
\AAA_0=\pt{\mat{cccc}{
0&m_-&n_1+1&m_+\\
m_-&0&m_+&n_2+1\\
n_1&m_+&0&m_-\\
m_+&n_2&m_-&0
}}\ ,
\end{eqnarray}
where
 $$m_\pm=(c\pm c')/4, \ \ \ \ \ \ n_1=(a-1)/2, \ \ \ \ \ \ n_2=(b-1)/2.$$
The symplectic eigenvalue $\nu$ in the  definition of the logarithmic negativity can be expressed in terms of the elements of these matrices as
 \begin{eqnarray}\label{nu}
 \nu&=&\pg{
\pq{
\frac{a+b}{2}-\sqrt{4\,m_-^2+\frac{(a-b)^2}{4}-4\pt{\frac{a-b}{a+b}}^2\,m_+^2}
\ }^2
%}\nn\\ &&\rpg{
-\frac{16\,a\,b}{(a+b)^2}\,m_+^2
}^{1/2}\ .
\end{eqnarray}
Correspondingly, these matrices can be used to determine an explicit expression for $E_S$
\begin{eqnarray}\label{ES0}
E_S=2\pq{
1+ \frac{2\,
n_1\,\xi_1^2+ 2\,n_2\,\xi_2^2+4\,\xi_1\xi_2\,m_-\cos\pt{\phi_1+\phi_2}
}{\xi_1^2+\xi_2^2}
}
\end{eqnarray}
that is minimized for
\begin{eqnarray}\label{minphixi}
\cos\pt{\phi_1+\phi_2}&=&-\frac{m_-}{\abs{m_-}}
\nn\\
\frac{\xi_1}{\xi_2}&=&\frac{2\abs{m_-}}{n_1-n_2+\sqrt{4\,m_-^2+\pt{n_1-n_2}^2}}
\end{eqnarray}
and the corresponding minimum is
\begin{eqnarray}\label{ES}
{\rm min}_{\phi_j,\xi_j}\, E_S=2\pq{1+n_1+n_2-\sqrt{4\ m_-^2+\pt{n_1-n_2}^2}}\ .
\end{eqnarray}
We note that 
if $m_+=0$ (i.e. $c=-c'$), then $2\nu$ is equal to Eq.~\rp{ES}, that is 
\begin{eqnarray}\label{nuES}
2\ \nu\Bigl|_{m_+=0}={\rm min}_{\phi_j,\xi_j}\, E_S\ .
\end{eqnarray}
This result is important because joins directly an entanglement measure, namely the logarithmic negativity, to the field observables, namely the variances of the field quadratures. As we have seen this is true only for the specific class of states for which $m_+=0$, that correspond to the condition $\av{b_j\ b_k\da}=0$ with $j\neq k=1,2$ (or equivalently $\av{x_1\al{0}\, x_2\al{0}}=-\av{x_1\al{\pi/2}\, x_2\al{\pi/2}}$). A related result has been previously discussed in Ref.~\cite{Adesso} where the symplectic eigenvalue has been shown to be equal to the EPR correlations in the case of symmetric states, for which $a=b$.

As discussed in the main text, the condition $m_+=0$ is relevant for the 
study of entanglement between the spectral components of stationary continuous fields.
The general, corresponding correlation matrix takes the form of Eq.~\rp{Ainfty3}, for which the correlations of the form $\av{b_j\,b_k\da}$ are zero. 
We note that in this case Eq.~\rp{ES0}  is equal to twice Eq.~\rp{DeltaXxi} when $m_-=\abs{m}$ and $\phi_1+\phi_2=\theta_1+\theta_2+{\rm arg}[m]$, where $\arg(m)$ is the phase of the complex parameter $m$, that is introduced in Eq.~\rp{Ainfty3}. The minimization of Eq.~\rp{DeltaXxi} is therefore similar to Eqs.~\rp{minphixi} and \rp{ES} (see Eq.~\rp{Smin} in the main text).
The corresponding covariance matrix is given by 
\begin{eqnarray}\label{barCC}
\overline \CC=\pt{\mat{cccc}{
2\,n_1+1&0&2\,{\rm Re}[m]&2\,{\rm Im}[m]\\
0&2\,n_1+1&2\,{\rm Im}[m]&-2\,{\rm Re}[m]\\
2\,{\rm Re}[m]&2\,{\rm Im}[m]&2\,n_2+1&0\\
2\,{\rm Im}[m]&-2\,{\rm Re}[m]&0&2\,n_2+1
}}\ ,
\end{eqnarray}
We note that this matrix  and the matrix in Eq.~\rp{Ainfty3} are not, in general, in the standard form described by Eqs.~\rp{standardC} and \rp{standardA}. They can be cast in standard form by means, for example, of the single mode rotation that perform the transformation $\overline a(\Omega)\to \ee^{-\ii\arg(m)}\overline a(\Omega)$. Thereby the resulting matrices are equal to Eqs.~\rp{standardC} and \rp{standardA} with $c'=-c$ and $m_+=0$. Therefore the corresponding minimum symplectic eigenvalue of the partially transposed covariance matrix has the form of Eq.~\rp{nu} with $m_-=\abs{m}$ and $m_+=0$, and it is explicitly given by Eqs.~\rp{nuOmega} and \rp{Smin}.

\section{The power spectrum of the stationary homo/hetero-dyne photocurrent}\label{apphomodyne-heterodyne}

In homodyne and heterodyne detection techniques~\cite{Collett,Lvovsky,Yuen78,Yuen83,Lvovsky14,Barbosa} the signal field is mixed on a 50:50 beam splitter with a strong monochromatic field at the frequency $\omega_{LO}$, the local oscillator. When the frequency of the local oscillator is equal to the carrier frequency, $\omega_L$, then one has homdyne detection. Heterodyne detection corresponds, instead, to finite detuning $\Delta=\omega_{LO}-\omega_L\neq0$.  The fields at the two output ports of the beam splitter are detected and the corresponding photo-currents are subtracted to end up with a classical electronic signal (which contains informations about a particular quadrature of the signal field)
\begin{eqnarray}
&&\II\al{\theta}(t)\propto\ee^{\ii\Delta t}\ \ee^{\ii\theta}\ \alpha(t) +\ee^{-\ii\Delta t}\ \ee^{-\ii\theta}\ \alpha^*(t)
\end{eqnarray}
where  $\alpha(t)$ is a classical random variable, $\theta$ is the phase of the local oscillator, and with 
$\Delta=0$ corresponding to homodyne detection. 
Informations about the spectral components of a detected stationary signal are provided by the power spectrum that quantifies the strength of the fluctuations at specific frequencies. We will refer to it as the homodyne  or heterodyne spectrum. It is given by
\begin{eqnarray}\label{GG0}
\GG\al{\theta}(\epsilon)=\int_{-\infty}^{\infty}\dd t\ \ee^{\ii\epsilon  t}\av{\II\al{\theta}(t)\,\II\al{\theta}(0)}
\end{eqnarray}
where, here, the angular brackets have to be intended as ensemble averages over many experimental runs, and
where, we use the fact that the photocurrent is a real stationary random process, for which the two-times correlation function depends only on the time difference and is symmetric $\av{\II\al{\theta}(t)\,\II\al{\theta}(t')}=\av{\II\al{\theta}(\pm\pt{t-t'})\,\II\al{\theta}(0)}$. In particular this imply that 
\begin{eqnarray}\label{appGsimm}
\GG\al{\theta}(\epsilon)=\GG\al{\theta}(-\epsilon)\ .
\end{eqnarray}
The power spectrum can be equivalently expressed trough the relation
\begin{eqnarray}\label{appGG00}
\av{\widetilde{\II\al{\theta}}(\epsilon)\,\widetilde {\II\al{\theta}}(\epsilon')}=\delta\pt{\epsilon+\epsilon'}\ \GG\al{\theta}(\epsilon)
\end{eqnarray}
where $\widetilde{ \II\al{\theta}}(\epsilon)=\frac{1}{\sqrt{2\pi}}\int_{-\infty}^{\infty}\dd t\ \ee^{\ii\omega t} \II\al{\theta}(t)$.
In practice the homo/hetero-dyne spectrum is evaluated in an approximate way by 
filtering the photocurrent with a filter function of length $\tau$ 
\begin{eqnarray}\label{appfilterIcurrent}
\overline {\II\al{\theta}_{\tau}}(\epsilon,t)&=&
\frac{1}{\sqrt{\tau}}\int_{t-\tau}^t \dd s\   \ee^{\ii\epsilon s} \II\al{\theta}(s)
\end{eqnarray}
and then calculating the corresponding autocorrelation function,
\begin{eqnarray}\label{appGG}
\GG\al{\theta}_\tau(\epsilon)&=&\av{\abs{\overline{\II\al{\theta}_{\tau}}(\epsilon,t)}^2}\ .
\end{eqnarray}
When $\tau$ is sufficiently large then the spectral properties of the stationary signal can be resolved and the power spectrum is well approximated
\begin{eqnarray}\label{applimGG}
 \lim_{\tau\to\infty}\GG\al{\theta}_\tau(\epsilon) =\GG\al{\theta}(\epsilon) \ .
\end{eqnarray}
This relation can be demonstrated as follows. The filtered photocurrent is equivalently given by 
\begin{eqnarray}
\overline{\II\al{\theta}_{\tau}}(\epsilon,t)&=&
\int_{-\infty}^{\infty}\dd\omega\ \ee^{-\ii(\omega-\epsilon)\, t}\ \widetilde \phi_\tau^{step}(\omega-\epsilon)\ \widetilde{\II\al{\theta}}(\omega)
\end{eqnarray}
where the filter function $\widetilde \phi_\tau^{step}(\omega)$ is defined in Eq.~\rp{phistep}.
Thereby we find
\begin{eqnarray}
&&\GG\al{\theta}_\tau(\epsilon)=
\int_{-\infty}^{\infty}\dd\omega\ \int_{-\infty}^{\infty}\dd\omega'\ \ee^{-\ii(\omega-\epsilon)\, t}\ \ee^{\ii(\omega'-\epsilon)\, t}\ 
%\nn\\&&\hspace{1.5cm}\times\ 
\widetilde \phi_\tau^{step}(\omega-\epsilon)\ 
 {\widetilde \phi_\tau^{step}(\omega'-\epsilon)}^*\
\av{\widetilde{\II\al{\theta}}(\omega)\, \widetilde{\II\al{\theta}}(-\omega')}
\nn\\
&&\hspace{1cm}=
\int_{-\infty}^{\infty}\dd\omega
\ \abs{\widetilde{\phi}_\tau^{step}(\omega-\epsilon)}^2\ \GG\al{\theta}(\omega)\ .
\end{eqnarray}
where we have used Eq.~\rp{appGG00}. In the large $\tau$ limit the modulus square of the filter function is equal to a delta function, obtaining therefore Eq.~\rp{applimGG}.

We remark that the filtered photocurrent  in Eq.~\rp{appfilterIcurrent} is complex and therefore is not directly related to a measurable (real) quantity. however we note that the same result presented in Eq.~\rp{applimGG} is obtained if, instead, we use the real photocurrent 
\begin{eqnarray}\label{appfilterIcurrent2}
{\JJ\al{\theta}_{\tau}}(\epsilon,t)&=&
\frac{1}{\sqrt{2\,\tau}}\int_{t-\tau}^t \dd s\  2\, \cos(\epsilon\,t+\varphi)\ \II\al{\theta}(s) \ .
\end{eqnarray}
 In particular the corresponding power spectrum is independent from the phase $\varphi$. In order to demonstrate this statement we rewrite 
Eq.~\rp{appfilterIcurrent2} as
\begin{eqnarray}
{\JJ\al{\theta}_{\tau}}(\epsilon,t)&=&
\frac{1}{\sqrt{2}}
\int_{-\infty}^{\infty}\dd\omega\ \pq{\ee^{\ii\varphi}\,
\ee^{-\ii(\omega-\epsilon)\, t}\ \widetilde \phi_\tau^{step}(\omega-\epsilon)
%}\nn\\&&\rpq{
+\ee^{-\ii\varphi}\,\ee^{-\ii(\omega+\epsilon)\, t}\ \widetilde \phi_\tau^{step}(\omega+\epsilon) 
}\ \widetilde{\II\al{\theta}}(\omega)\ .
\end{eqnarray}
Hence the corresponding autocorrelation function is given by
\begin{eqnarray}
\GG\al{\theta}_\tau(\epsilon)&=&\frac{1}{2}
\int_{-\infty}^{\infty}\dd\omega\ 
\lpq{
\abs{\widetilde \phi_\tau^{step}(\omega-\epsilon)}^2
+
\abs{\widetilde \phi_\tau^{step}(\omega+\epsilon) }^2
}
\nn\\&&\rpq{
+
\ee^{2\ii(\epsilon\, t+\varphi)}\ \widetilde \phi_\tau^{step}(\omega-\epsilon)
\,\widetilde \phi_\tau^{step}(-\omega-\epsilon)
%\nn\\&&\rpq{
+
\ee^{-2\ii\,(\epsilon\, t+\varphi)}\ \widetilde \phi_\tau^{step}(\omega+\epsilon) 
\widetilde \phi_\tau^{step}(-\omega+\epsilon) 
}
\GG\al{\theta}\pt{\omega}\nn\\
\end{eqnarray}
where we have used the relation ${\widetilde \phi_\tau^{step}(\omega)}^*=\widetilde \phi_\tau^{step}(-\omega)$ and Eq.~\rp{appGG00}. Finally using Eq.~\rp{rel-filter} and Eq.~\rp{appGsimm} we find that, in the limit of large $\tau$, this equation reduces to Eq.~\rp{applimGG}.
 
The photocurrent is directly related to the properties of the detected field. In fact, the ensemble average in Eq.~\rp{GG0} can be equivalently interpreted as a quantum average over an operator of the form 
\begin{eqnarray}\label{It}
I_\Delta\al{\theta}(t)= \ee^{\ii\,\Delta\,t}\ee^{\ii\theta}\ a(t) +\  \ee^{-\ii\,\Delta\,t}\ee^{-\ii\theta}\ a\da(t) 
\end{eqnarray}
where now $a(t)$ and $a\da(t)$ are quantum operators for the detected field, and  the results discussed above, in terms of classical photocurrents, can be straightforwardly rephrased in terms of this quantum operator.

Furthermore, while the results for the homodyne and heterodyne spectra are independent from the form of the filtered photocurrent, either Eq.~\rp{appfilterIcurrent} or Eq.~\rp{appfilterIcurrent2}, the choice of Eq.~\rp{appfilterIcurrent2}  is physically motivated by the fact that it results in a real filtered photocurrent that corresponds to an hermitian quantum operator, and hence it makes transparent the relation between the spectral properties of the detected photocurrent and the corresponding quantum observables of the stationary field. In particular the filtered photocurrent can be described by the hermitian operator
$$
 {J_{\Delta,\tau}\al{\theta}}(\epsilon)=\frac{N_\tau}{\sqrt{2\tau}}\int_{t-\tau}^t\dd s\ \cos\pt{\epsilon\, s+\varphi}\ I\al{\theta}(s)
$$ 
where the normalization factor $N_\tau$ is appropriately chosen in order to satisfy the commutation relation for quadrature operators $\pq{ {J\al{\theta}_{\Delta,\tau}}(\epsilon,t), {J\al{\theta+\frac{\pi}{2}}_{\Delta,\tau}}(\epsilon,t)}=2\,\ii$, namely $N_\tau=\sqrt{(1+\tau\,\epsilon)/(2+\tau\,\epsilon)}$. 
Thus, the filtered photocurrent can be expressed as the sum of two filtered quadrature operators for the 
two frequency bands of width $1/\tau$ each, centred at the frequencies  $\Delta\pm\epsilon$,
\begin{eqnarray}\label{hatI}
 {J_{\Delta,\tau}\al{\theta}}(\epsilon,t)=\frac{N_\tau}{\sqrt{2}}\pq{\overline{x_{\tau}\al{\theta+\varphi}}(\Delta+\epsilon,t)+ \overline{x_{\tau}\al{\theta-\varphi}}(\Delta-\epsilon,t)}
\end{eqnarray}
where 
\begin{eqnarray}\label{Itauepsilon}
\overline{x_{\tau}\al{\theta}}(\Delta\pm\epsilon,t)=\ee^{\ii\theta}\ \overline {a_\tau}\pt{\Delta\pm\epsilon,t}+\ee^{-\ii\theta}\ \overline {a_\tau\da}\pt{-\Delta\mp\epsilon,t}
\end{eqnarray}
with the annihilation operator, of a single band filtered mode, defined as in Eq.~\rp{bara}. In the limit of large $\tau$, Eq.~\rp{hatI} reduces to Eq.~\rp{narrowI} and Eq.~\rp{narrowIhetero}, when, respectively, $\Delta=0$ and $\Delta\neq0$.

\section{A single-mode cavity with a membrane in the middle: Input-otput theory}\label{inputoutput}

We consider a single-mode Fabry-Perot cavity with a membrane in the middle as discussed in the main text. The quantum Langevin equations~\cite{Gardiner} for the creation and annihilation operators of a cavity photon $a\da,a$ and of a membrane phonon $b\da,b$, in the linearized regime~\cite{Genes}, can be expressed in matrix form as
\begin{eqnarray}\label{QLE}
\dot\va(t)=\MM\ \va(t)+\QQ\ \va_{in}(t)
\end{eqnarray}
where $\va$ is the column vector of system operators $\va=\pt{a,b,a\da,b\da}^T$, the matrix of coefficients $\MM$ is given by
\begin{eqnarray}
\MM=
\pt{
\begin{array}{cccc}
-\kappa_1-\kappa_2-\ii\delta  &-\ii g   & 0 &-\ii g  \\
 -\ii g & -\gamma-\ii{\omega_m}  &-\ii g&0   \\
 0 &  \ii g & - \kappa_1-\kappa_2+\ii\delta &\ii g\\
 \ii g& 0 &\ii g & -\gamma+\ii{\omega_m}
\end{array}
}\ , 
\end{eqnarray}
with the parameters defined in the main text, and $\va_{in}$ is the vector of input noise operators $\va_{in}=\pt{a_1^{in},a_2^{in},b^{in},{a_1^{in}}\da,{a_2^{in}}\da,{b^{in}}\da}^T$, which includes the two inputs of the cavity corresponding to the two mirrors; Finally $\QQ$ is the $4\times 6$ matrix
\begin{eqnarray}
\QQ=\pt{
\begin{array}{cccccc}
\sqrt{2\kappa_1}&\sqrt{2\kappa_2}&0&0&0&0\\
0&0&\sqrt{2\gamma}&0&0&0\\
0&0&0&\sqrt{2\kappa_1}&\sqrt{2\kappa_2}&0\\
0&0&0&0&0&\sqrt{2\gamma}\\
\end{array}
}\ .
\end{eqnarray}
In general the system dynamics can be divided into two main parameter regimes~\cite{Genes}. When the real part of all the eigenvalues of the matrix $\MM$ is negative then the system is stable and approaches a steady state at large times. If, on the other hand, some eigenvalues have a positive real part then the system is not stable, the populations of the modes explode and no steady state is reached. In this second case the linearized model in Eq.~\rp{QLE} is not a valid description of the optomechanical dynamics.
All the results presented in the main text correspond to the regime of opomechanical stability.

The steady state corresponding to Eq.~\rp{QLE} can be easily obtained in Fourier space. We introduce the Fourier transformed operators  $\tilde\va(\omega)=\frac{1}{\sqrt{2}}\int\dd t\ee^{\ii\omega t}\va(t)$, hence
\begin{eqnarray}\label{aomega}
\tilde\va(\omega)=-\pt{\MM+\ii\omega}^{-1}\QQ\ \tilde\va^{in}(\omega)\ .
\end{eqnarray}
We are interested in the field leaking out by the two cavity mirrors. According to the input output theory~\cite{Gardiner}, the operators for the output fields can be expressed in terms of the system and of  the input noise operators as $a^{out}_j=\sqrt{2\kappa_j}\ a-a^{in}_j$, where $j=1,2$ distinguish the two output channels corresponding to the two cavity mirrors. 
In order to express these relations in matrix form we introduce the $4\times 6$ matrix 
\begin{eqnarray}
\ZZ=\pt{
\begin{array}{cccccc}
1&0&0&0&0&0\\
0&1&0&0&0&0\\
0&0&0&1&0&0\\
0&0&0&0&1&0\\
\end{array}
}
\end{eqnarray}
that when applied to the vector of input operators gives $\ZZ\, \va_{in}=\pt{a^{in}_1,a^{in}_2,{a^{in}_1}\da,{a^{in}_2}\da}^T$, and selects only the noise operators corresponding to the two output channels.
Thus, the vector of output operators $\va_{out}=\pt{a^{out}_1,a^{out}_2,{a^{out}_1}\da,{a^{out}_2}\da}^T$ can be written as
 \begin{eqnarray}
\va_{out}
=\ZZ\,\QQ^T\ \va-\ZZ\ \va_{in}\ .
\end{eqnarray}
Using Eq.~\rp{aomega} we find
\begin{eqnarray}
\tilde\va_{out}(\omega)=-\ZZ\pq{\QQ^T\, \pt{\MM+\ii\omega}^{-1}\QQ\ +\id}\tilde\va_{in}(\omega)\ ,
\end{eqnarray}
and the corresponding power spectrum matrix is
\begin{eqnarray}
\widetilde\PP_{out}(\omega)&=&\ZZ\pq{\QQ^T\, \pt{\MM+\ii\omega}^{-1}\QQ\ +\id}\CC_{in}
%\nn\\&&\hspace{1.2cm}\times
\pq{\QQ^T\, \pt{\MM-\ii\omega}^{-1}\QQ\ +\id}\ZZ^T
\end{eqnarray}
where
$\CC_{in}$ is the correlation matrix of the input noise operators defined as $\delta(\omega+\omega')\, \CC_{in}=\av{\tilde\va_{in}(\omega)\ {\tilde\va_{in}(\omega)}^T}$, and it is given by
\begin{eqnarray}
C_{in}=\pt{
\begin{array}{cccccc}
0&0&0&1&0&0\\
0&0&0&0&1&0\\
0&0&0&0&0&n_T+1\\
0&0&0&0&0&0\\
0&0&0&0&0&0\\
0&0&n_T&0&0&0
\end{array}
}
\end{eqnarray}
where $n_T$ is the number of thermal excitations of the mechanical oscillator.
The explicit result for $\widetilde\PP_{out}(\omega)$ is given in Eq.~\rp{AAAout}.

\end{document}